\documentclass[preprint,onecolumn,numberedappendix,appendixfloats]{emulateapj}
\usepackage{natbib}
\usepackage{amsmath,amssymb}
\usepackage{wasysym}
\usepackage{graphicx,epsf}
\usepackage{subfigure}

\def\TWOFIGs#1#2#3#4{
	\begin{figure*}[htbp] 
	\centering
		\subfigure[#3]{
	         \includegraphics[width=0.45\hsize]
	         {#1/spec/histo.peak#2.ps}\label{fig:SpecHistogram#1#2}}\quad
		\subfigure[#4]{
		\includegraphics[bb= 0 0 800 600, clip, viewport= 50 0 490 420, width=0.45\hsize]         
	         {#1/spec/#1#2.eps}\label{fig:SpecDistribution#1#2}}
	         \caption{Peak #1#2}\label{fig:twofig#1#2}
	\end{figure*}
}

\begin{document}

\title{Reducing Systematic Error in Cluster Scale Weak Lensing}\thanks{Based in part on data collected at Subaru Telescope and obtained from the SMOKA, which is operated by the Astronomy Data Center, National Astronomical Observatory of Japan.}
\author{
Yousuke Utsumi\altaffilmark{1,2},
Satoshi Miyazaki\altaffilmark{1,3},
Margaret J. Geller\altaffilmark{4},
Ian P. Dell'Antonio\altaffilmark{5},
Masamune Oguri\altaffilmark{6},
Michael J. Kurtz\altaffilmark{4},
Takashi Hamana\altaffilmark{1}, and
Daniel G. Fabricant\altaffilmark{4}
}
\email{yutsumi@naoj.org}

\altaffiltext{1}{National Astronomical Observatory of Japan, 2-21-1 Osawa,  Mitaka, Tokyo 181-8588, Japan}
\altaffiltext{2}{Hiroshima Astrophysical Science Center, Hiroshima University, 1-3-1 Kagamiyama, Higashi-Hiroshima, Hiroshima 739-8526, Japan}
\altaffiltext{3}{The Graduate University for Advanced Studies, 2-21-1 Osawa, Mitaka, Tokyo 181-8588, Japan}
\altaffiltext{4}{Smithsonian Astrophysical Observatory, 60 Garden St., Cambridge, MA 02138, USA}
\altaffiltext{5}{Department of Physics, Brown University, Box 1843, Providence, RI 02912, USA}
\altaffiltext{6}{Kavli Institute for the Physics and Mathematics of the Universe (Kavli IPMU, WPI), University of Tokyo, 5-1-5 Kashiwanoha, Kashiwa, Chiba 277-8583, Japan}

\begin{abstract}
Weak lensing provides an important route toward collecting samples of clusters of galaxies selected by mass. Subtle systematic errors in image reduction can compromise the power of this technique.
We use the B-mode signal to quantify this systematic error and to test methods for reducing this error.
We show that two procedures are efficient in suppressing systematic error in the B-mode: (1)
refinement of the mosaic CCD warping procedure to conform to absolute celestial coordinates
and (2)  truncation of the smoothing procedure on a scale of 10$^{\prime}$.
Application of these procedures reduces the systematic error to 20\% of its original amplitude.
We provide an analytic expression for the distribution of the highest peaks in noise maps that can be used to estimate the fraction of false peaks in the weak lensing $\kappa$-S/N maps as a function of the detection threshold. Based on this analysis we select a threshold S/N = 4.56 for identifying an uncontaminated set of weak lensing peaks in two test fields covering a total area of $\sim 3$deg$^2$. Taken together these fields contain seven peaks above the threshold. Among these, six are probable systems of galaxies and one is a superposition. We confirm the reliability of these peaks with dense redshift surveys, x-ray and imaging observations. 
The systematic error reduction procedures we apply are general and can be applied to future large-area weak lensing surveys. Our high peak analysis suggests that with a S/N threshold of 4.5, there should be only 2.7 spurious weak lensing peaks even in an area of 1000 deg$^2$ where we expect $\sim$ 2000
peaks based on our Subaru fields.

\end{abstract}

\section {Introduction}
Weak lensing is a fundamental tool of modern cosmology. A weak-lensing map provides a weighted ``picture'' of the projected surface mass density and thus a route to identifying clusters of galaxies
selected by mass \citep[e.g.][]{2002ApJ...580L..97M,2003ApJ...591..662D,2003A&A...407..869S,2005A&A...442...43H,2006ApJ...643..128W,2007A&A...462..459G,2007ApJ...669..714M,2012ApJ...748...56S}. Subtle issues limit the application of weak lensing maps as sources of
cluster catalogs. A serious astrophysical limitation is the projection of large-scale structure along the line-of-sight \cite[see e.g.][]{2002ApJ...575..640W,2004MNRAS.350..893H,2012MNRAS.425.2287H}. An observational limitation is the presence of systematic errors in the maps \citep{2010ApJ...709..832G,2012ApJ...750..168K} . Here we focus on a method of reducing these errors.

\citet{2001ApJ...557L..89W} and \citet{2002ApJ...580L..97M} first identified rich clusters from weak lensing maps.
Since then larger and larger weak lensing surveys with increasingly sophisticated reduction techniques
have led to a steep increase in the number of clusters initially detected in weak lensing maps
\citep{2005A&A...442...43H,2006ApJ...643..128W,2007A&A...462..875S,2007A&A...462..459G,2007ApJ...669..714M,2007A&A...462..473M,2008MNRAS.385..695B,2007A&A...470..821D,2011MNRAS.413.1145B,2012ApJ...748...56S}. \citet{2012ApJ...748...56S} used the CFHT to  carry out the largest area survey to date covering 64 deg$^2$. Among the 301 weak lensing peaks with signal-to-noise ratio greater than 3.5, only 126 have a corresponding brightest cluster galaxy within
3$^{\prime}$ of the weak lensing peak. \citet{2012ApJ...748...56S} conclude that many of the weak lensing peaks, even above this threshold, are probably noise.

In a detailed reanalysis of the 2.1deg$^2$ GTO field of \cite{2002ApJ...580L..97M,2007ApJ...669..714M,2012ApJ...750..168K} show that the highest peak of the B-mode map (in this map the sources ellipticities are rotated by 45$^\circ$ and there should be no signal) has signal-to-noise, S/N $\sim 4.25$
and there are four peaks in the B-mode map with S/N$> 3.7$. It is interesting that \cite{2012ApJ...748...56S} find a similar number of high-significance peaks per unit area in their B-mode maps. Analysis of the B-mode map prompted \cite{2012ApJ...750..168K} to set their threshold for weak lensing cluster detection at 4.25.
Above this threshold 2/3 of the weak lensing peaks correspond to individual massive clusters.

Here we investigate possible underlying sources of the high peaks in the B-mode maps. Our goal is reduction of the amplitude and  frequency of these peaks as a basis for construction of more robust catalogs of weak lensing cluster detections. We apply our revised procedures to two fields observed with the Subaru Telescope: the GTO field of \cite{2002ApJ...580L..97M,2007ApJ...669..714M} and a small portion of the DLS F2 field
reobserved with Subaru.

In Section 2, we review the standard analysis of the Subaru imaging data. In Section 3 we review the construction of the galaxy catalog. In Section 4 we describe the basic weak lensing analysis we use.
In Section 5 we develop a revised procedure for reducing systematics in the  B-mode map. We demonstrate that there is excess power in the B-mode map on small and large scales relative to noise maps. Two procedures are effective in suppressing systematic in the B-mode map: (1) warping the image onto absolute sky coordinates and (2) introducing a large-scale (10$^{\prime}$) cutoff in the smoothing kernel applied to
construct the weak lensing map. We define the weak lensing peaks and investigate thresholds in Section 6. In Section 7 we note some differences in our analysis of the GTO and DLS Subaru observations.
Section 8 provides an analytic expression for the probability of finding the highest peak in the noise maps as spurious peaks in a weak lensing map. In Section 9 we test the lensing maps of our two fields by comparing the peaks with systems identified in dense redshift surveys. We demonstrate that a threshold S/N  = 4.56 produces a robust catalog of peaks with few false positives.

Unless otherwise stated,
we adopt the standard concordance cosmology $(h=0.72, \Omega_{m}=0.27, \Omega_{\Lambda}=0.73)$
and the AB magnitude system  throughout this paper.

\section{Imaging Data \& Image Reduction}

In this section we describe a revised reduction procedure that reduces systematic errors in weak lensing maps.
As examples of the revised procedure, 
we use two regions  imaged with Suprime-Cam on the Subaru Telescope \citep{2002PASJ...54..833M}.
The GTO field covers $\sim 2$ deg$^2$,
centered on (16:04, +43:12) and was observed by \citet{2002ApJ...580L..97M}. We re-observed 
the western $\sim 1$ deg$^2$ portion, (9:16, +30:00), of the
Deep Lens Survey F2 field \citep[DLS F2][]{2002SPIE.4836...73W}.
The new observations were taken in 0.75$^{\prime\prime}$ seeing.
The GTO field consists of 9 Suprime-Cam pointings; the DLS observations consist of  4 Suprime-Cam pointings.
Table \ref{GTOframelist} shows  the numbering scheme for Subaru pointings in both the GTO and DLS fields. 

We retrieved data from the SMOKA data archive server \citep{2002ASPC..281..298B}.
These data satisfy the condition  FILTER=W-C-RC and
$\rm{PSF\_SIGMA}<0.9$ to obtain good image quality.
$\rm{PSF\_SIGMA}$ is the typical FWHM (arcsec) of stellar images
evaluated by SMOKA using SExtractor \citep{1996A&AS..117..393B}.
If the total exposure exceeds 2,000 seconds,
we select frames with the smallest PSF\_SIGMA 
and reject  frames with larger PSF\_SIGMA 
even if PSF\_SIGMA is $\lesssim 0.9 ^{\prime\prime}$.
As a result, we reject about 6 of 11, 10 of 15 and 1 of 5 exposures
for GTO\_0, GTO\_4, and GTO\_5, respectively.
Table \ref{GTOframelist} gives the total exposure times, observation dates, filters, typical seeing, and effective limiting magnitude for each subfield.
Imaging observation for the DLS F2 field were made with Suprime-Cam in January 1, 2008.
Each of the 4 subfields is a 3600 sec exposures taken
under stable seeing conditions. Table \ref{GTOframelist}
lists the FWHM of the PSF on stacked images in each of the four subfields.

\begin{table*}[tdp]
\caption{Observation Log.}
\begin{center}
\begin{tabular}{c|cc|c|c|c|c|c}
Name	&	Ra	&	Dec	&	Exp. time (sec)	&	Observation date	&	Filter	&	Seeing	&	Limit mag\\
\hline
\hline
GTO\_0 &	16:04:44.5	&	+43:11:12	&	2100	&	2001/4/23, 25	&	$R_{\rm C}$	&	0.66	&	25.38\\
GTO\_1 &	16:07:39.0	&	+43:12:19	&	2400	&	2001/4/23, 25	&	$R_{\rm C}$	&	0.77	&	25.28\\
GTO\_2 & 	16:07:39.0	&	+43:37:37	&	1800	&	2001/4/23	&	$R_{\rm C}$	&	0.72	&	25.44\\
GTO\_3 &	16:04:44.6	&	+43:36:30	&	1800	&	2001/4/24	&	$R_{\rm C}$	&	0.71	&	25.24\\
GTO\_4 &	16:01:46.8	&	+43:37:37	&	2250	&	2001/4/24, 25, 5/18	&	$R_{\rm C}$	&	0.65	&	25.17\\
GTO\_5 &	16:01:46.8	&	+43:12:19	&	2250	&	2001/4/24, 25	&	$R_{\rm C}$	&	0.70	&	24.90\\
GTO\_6 &	16:01:46.8	&	+42:47:01	&	1800	&	2001/4/24, 25	&	$R_{\rm C}$	&	0.66	&	25.19\\
GTO\_7 &	16:04:43.0	&	+42:47:01	&	1800	&	2001/4/25	&	$R_{\rm C}$	&	0.65	&	25.18\\
GTO\_8 &	16:07:39.0	&	+42:47:01	&	1800	&	2001/4/25	&	$R_{\rm C}$	&	0.64	&	25.32\\
\hline
DLSF2\_f0	&	09:16:30.9    &	+29:17:40	&	3600	&	2008/1/8	&	$i'$	&	0.79	&	25.50\\
DLSF2\_f1	&	09:16:31.0    &	+29:39:40	&	3600	&	2008/1/8	&	$i'$	&	0.72	&	25.52\\
DLSF2\_f2	&	09:16:31.0    &	+30:01:40	&	3600	&	2008/1/8	&	$i'$	&	0.72	&	25.33\\
DLSF2\_f3	&	09:16:31.0    &	+30:23:41	&	3600	&	2008/1/8	&	$i'$	&	0.76	&	25.51\\
\hline
\end{tabular}
\end{center}
\footnotetext{The GTO pointings  fill a $3\times3$ square. They are numbered so that the central field is 0 and numbers go from 1 in the central left (east) field  clockwise to 8 in the southeast corner. The DLS pointings fill a $1\times4$ rectangle: 0 is the southernmost pointing and the numbers increase to the north. For the GTO field the numbering is the same as in \citet{2012ApJ...750..168K}}
\footnotetext{Detail of the limiting magnitude as described in Section \ref{galaxycatalog}.}
\label{GTOframelist}
\end{table*}%

The basic image reduction follows 
\citet{2002ApJ...580L..97M,2007ApJ...669..714M} and \citet{2012ApJ...750..168K}.
However, we make one major revision: we use independent astrometric data to 
register the images accurately with respect to  the celestial coordinate
while avoiding the  introduction of  any additional image warp.

The preliminary reduction of the raw image on each CCDs is standard.
We use SExtractor to identify objects on the
overscan-subtracted, flat-fielded, sky-subtracted image.
We identify stellar objects from their easily identifiable linear size-magnitude relation.
We  manually select the stellar object catalog using this relation.

After completing the preliminary reduction, there are three additional steps in obtaining the final mosaic of stacked images;  one of the steps is new.
First we determine the accurate location of each CCD in instrument coordinates (Section \ref{step1}).
Second, we apply the image alignment warping procedure of \citet{2002ApJ...580L..97M,2007ApJ...669..714M}
(Section \ref{originalwarp}). Finally 
we introduce a new additional warp ``Registration onto the celestial coordinate'' (Section \ref{newwarp}).

\subsection{ Solving the Basic Mosaic Geometry}
\label{step1}

We parameterize the accurate location of each CCD in instrument coordinates 
by the displacement and the rotation of each CCD $(\Delta x, \Delta y, \Delta \phi)_{\rm c}$,
the telescope pointing offset between dithered exposures $(\Delta X, \Delta Y, \Delta \Phi)_{\rm e}$,
and  the optical distortion of the wide field corrector lens.
The distortion parameters are well modeled by a 4th-order polynomial function:
\begin{eqnarray}
	\frac{R-r}{r}=a r + br^{2}+cr^{3}+dr^{4}
\end{eqnarray}
where $R$ and $r$ are distances from the optical axis in units of pixels on the CCD.
$R$ is the original instrument coordinate and $r$ is the distortion-free coordinate.  

To construct a mosaic we need a
flux calibration to correct for the differences in sensitivity from chip to chip ($\Delta f_{c}$).
These differences originate from differences in amplifier gain,  differences in the normalization of flat fielding, and
from corrections for the transmission of the atmosphere from exposure to exposure $\Delta f_{e}$.
We derive this calibration by comparing the fluxes for unsaturated stars in overlapping regions of the frames.
We call these parameters the ``mosaicking rule''.

Operationally, we reduce the rotational component ($\Delta\phi, \Delta\Phi$)
to the linearized form of the rotational matrix consisting of four components ( ($\phi_{ij}$), ($\Phi_{ij}$))
where $i,j \in (1,2)$.
Because this matrix can treat not only rotation but also expansion, we include the first coefficient 
of the distortion parameter in this matrix.
Varying the basic mosaic geometry parameters $(\Delta x, \Delta y, \phi_{ij}, \Delta f)_{\rm c}$,
$(\Delta X, \Delta Y,  \Phi_{ij}, \Delta f)_{\rm e}$, ($b, c, d $),
we minimize 
\begin{eqnarray}
	\Delta^{2} = \sum_{e>f }\sum_{i}^{\rm all~stars}
	\left(
	\frac{(\vec{\eta}_{i}^{(e)}-\vec{\eta}_{i}^{(f)})^{2}}{\sigma_{\eta}^{2}}
	+ \frac{(f_{i}^{(e)}-f_{i}^{(f)})^{2}}{\sigma_{f}^{2}}
	\right)
\end{eqnarray}
where $\vec{\eta}_{i}^{(e)}$ and $f_{i}^{(e)}$ are the
$e$-th exposure's position in distortion-free coordinates and the corrected flux for $i$-th unsaturated star and $\sigma_{i},~i\in (\eta, f)$ are the typical error in the measurements of position and flux, respectively.
We fix parameters for the first exposure of the series and the left bottom chip, i.e. 
$(\Delta x, \Delta y, \Delta \phi, \Delta f)_{\rm 0}=\vec{0}$,
$(\Delta X, \Delta Y, \Delta \Phi, \Delta f)_{\rm 0}=\vec{0}$
as constraints.

This mosaicking rule is calculated by \emph{imcat} \citep{1995ApJ...449..460K}.
The best fit parameters are obtained by minimizing the positional difference
of unsaturated control stars ($70\sim100$ stars per CCD) for each exposure.

The residual alignment error in this procedure is $\sim 0.5$ pixel rms (0.1$^{\prime\prime}$),
still large compared with typical distortions of the sources in weak lensing.
Many stacking pipelines adopt this ``typical stacking'' warp only.
We note that a typical stellar size is 0.6$^{\prime\prime}$; it should thus be possible to
align the stellar positions to  $\lesssim 0.06^{\prime\prime}$, one tenth of the stellar size.
The accuracy of the ``typical stacking'' is insufficient and probably below the level that can be obtained from the data.

\subsection{ Jelly Focal Plane Warp}\label{originalwarp}

Residual positional differences from exposure to exposure in the typical stacking procedure in Section \ref{step1} may result from non-symmetric features
of optical distortion (possibly due to  imperfect optical alignment)
not considered in the model.
Differential atmospheric dispersion is another source of asymmetry.

We parameterize the residuals relative to one another by a polynomial function of the field position; we expect the residuals to be continuous over the field of view. The residuals are
\begin{eqnarray}
	\Delta \xi=\sum_{l=0}^{6}\sum_{m=0}^{l}a_{lm,e}\xi^{l-m}\eta^{m},\quad
	\Delta \eta=\sum_{l=0}^{6}\sum_{m=0}^{l}b_{lm,e}\xi^{l-m}\eta^{m}\label{eq:pos}
\end{eqnarray}
where $\vec{\eta}=(\xi, \eta)$. We then obtain
the coefficients $a_{lm,e}$ and $b_{lm,e}$  by minimizing the variance of the residuals.
Each individual CCD image is ``warped'' using this polynomial correction prior to the stacking.
This process reduces the alignment error to $\sim 0.05$ pixel (0.014$^{\prime\prime}$).
\citet{2002ApJ...580L..97M, 2007ApJ...669..714M} and \citet{2012ApJ...750..168K} used this transformation
following \citet{1998astro.ph..9268K}.
We refer to this transformation as ``the original warp'',
and we call the results of this warp ``the original stacking''. The result  is similar to
\citet{2002ApJ...580L..97M,2007ApJ...669..714M} and \citet{2012ApJ...750..168K}.

\subsection{Registration onto the Celestial Coordinate}\label{newwarp}

The third transformation is new and we apply it as the final step in the reduction. 
The result of Section \ref{originalwarp} 
is that all exposures except for  the first one are registered
on the first exposure with an accuracy of $\sim0.05$ pixel, the tolerance we seek.
However, there is no guarantee that the geometry of the first exposure
is well matched to the celestial coordinate.

In practice, the absolute rotation of the instrument rotator is less well-known than its rotational angle relative to its zero point.
There is  no mechanical system to determine the zero point rotational angle for the
rotator accurately for Suprime-Cam; the rotational angle relative to the origin is, however, very well known. The possibility of this zero point offset suggests that the final Suprime-Cam image will be rotated relative to the absolute celestial coordinate.

To align the final image with the celestial coordinate frame (not some artificial distorted frame, but a physical frame),
we use an external catalog, USNO-A2.0 \citep{1998AAS...19312003M}.
This astrometric standard catalog has an accuracy of 0.4 arcsec
although USNO-A2.0 is not deep enough to match the Suprime-Cam catalogs.
The overlapping magnitude range includes moderately bright stars;
these stars are saturated but  do not have  blooming pixels.
It is thus difficult to measure their total intensity,
but it is possible to measure their position because they have circular symmetry.
We do not use the SDSS star catalog because it 
does not cover the enough of the sky to allow us to 
to extend our procedure  outside the GTO field.
We confirm that using the SDSS external astrometric catalog makes no significant difference in the warp relative to use of the USNO-A2.0.

In the same manner as the transformation in Section \ref{originalwarp}, we parameterise 
the residuals ($\Delta x, \Delta y$) of bright stars (they are
slightly separated from the linear stellar sequence
but have a comparable positional determination to the unsaturated stars
provided there are no  blooming pixels)
in the first exposure with respect to the USNO-A2.0 celestial coordinate
as a polynomial function of field position $\vec{x}=(x,y)$ (Equation \ref{eq:pos}).
This procedure  yields absolute astrometric object  positions accurate to $\sim 0.2^{\prime\prime}$.
The internal relative alignment of stars remains at the level of $\sim 0.05$ pixel.
We refer to this transformation  as ``the new warp''
and we call the result of this warp ``the new stacking''.

\begin{figure*}[htbp] 
   \centering
   \includegraphics[angle=-90,width=\hsize]{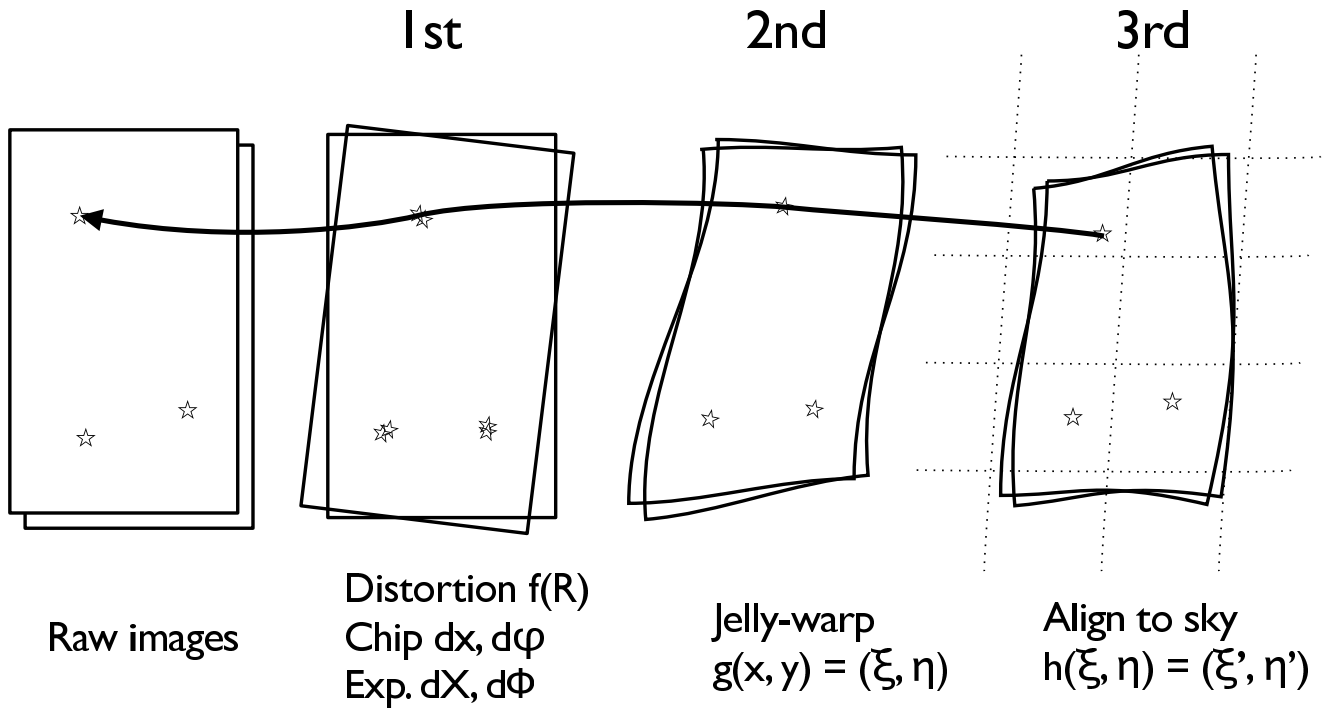} 
   \caption{Schematic of the three steps in the mosaic warp}
   \label{fig:cartoon}
\end{figure*}

We present a schematic picture to describe these three steps of the
image reduction in Figure \ref{fig:cartoon} and
summarize the three step image reduction pipeline (Section \ref {step1}, \ref{originalwarp}) compared to original one 
and typical one in Table \ref{tab:warprule}.
\begin{table}[htdp]
\caption{Summary of the warp rules}
\begin{center}
\begin{tabular}{c|ccc|c}
	&	New 		&	Original 	&	Typical 	&	Registration\\
\hline
\hline
Distortion f(R)	&	yes	&	yes	&	yes		&	$\sim0.5$ pixel	\\
Jelly-warp		&	yes	&	yes	&	no	&	$\sim0.05$ pixel	\\
Align to sky	&	yes	&	no	&	no	&	$\sim0.05$ pixel	\\
\hline
Astrometry	&	$\sim0.2$ arcsec	&	uncontrolled	&	depending on pipelines	&\\
\end{tabular}
\end{center}
\label{tab:warprule}
\end{table}%
Once we have all of the  parameters that describe these three steps,
we carry out the image warp all at once rather than applying the transformation step by step.
This integration of the procedures is necessary to minimize the artifacts due to the interpolation of pixels. We estimate
pixel values  via 3rd order bi-polynomial interpolation of $4\times4$ source pixels.
We employ the arithmetic mean for stacking.

\section{Galaxy Catalog}\label{galaxycatalog}

To construct the galaxy catalog from the mosaic image, we use the \emph{imcat} tools
``\emph{hfindpeaks}'' (object detection) and ``\emph{apphot}'' (photometry).
We use the half-light radius ``$r_{h}$'' as the size of an object.
Using this size, we can distinguish
galaxies from stars by requiring that $r_{h}$ for galaxies satisfy the relation:
\begin{eqnarray}
	r_{h} > r_{h}^{*}+\sigma_{r_{h}^{*}},
\end{eqnarray}
where $r_{h}^{*}$ and $\sigma_{r_{h}^{*}}$ are the half-light radius
of a a stellar image and its rms, respectively.

We merge our catalog of stars with the SDSS star catalog
to calibrate the photometric zeropoint (see. Equations \ref{photometric-calibration}).
Typically, the number density of stars to the depth of SDSS is several arcmin$^{-2}$;
the precision of the astrometric catalog is $\lesssim 1^{\prime\prime}$.

We determine the photometric zeropoints for each  subfield from the
the SDSS photometric catalog following \citet{2010ApJ...721.1680U}.
The transformation equation between the photometric systems is:
\begin{eqnarray}
Rc_{Subaru}-r'_{SDSS} &=& -0.314(r'_{SDSS}-i'_{SDSS})-0.011
\quad {\rm for} \quad r'_{SDSS}-i'_{SDSS} < 0.5\nonumber\\
i'_{Subaru}-i'_{SDSS} &=& 0.102(i'_{SDSS}-r'_{SDSS}) + 0.002
\quad{\rm for} \quad -0.5< i'_{SDSS}-r'_{SDSS}\label{photometric-calibration}
\end{eqnarray}

To assess the depth of our images,
we follow the standard procedure of SDFRED
to derive the limiting magnitude \citep{2002AJ....123...66Y,2004ApJ...611..660O}.
We give the $5 \sigma$ limiting magnitude in a 2$^{\prime\prime}$ photometric aperture  in Table \ref{GTOframelist}.

To remove low signal-to-noise regions around sufficiently bright stars, we need a 
procedure for constructing masks.
Stars brighter than 20 th magnitude are usually saturated
in a Suprime-Cam typical exposure with $\sim100$ sec integration.
We measure the actual PSF of Suprime-Cam over a wide dynamic range
by compiling the PSF of faint and bright stellar objects.
We divide the stellar samples into subsamples segregated by R-band magnitude, faint (20-20.5 magnitude), intermediate (18-18.5 magnitude) and bright (13.75-14.25 magnitude),
combine them by taking the arithmetic mean,
and then measure their radial profile.
Although we take a mean profile of bright stars,
we also reject exceedingly bright pixels by choosing an appropriate threshold for each images
in order to avoid saturated areas or blooming regions.

For stars with R $< 14$, 
there are not enough stars to determine the mean PSF.
Thus we consider an optical model of Suprime-Cam to determine the size of the surrounding halo.
The halo surrounding a bright star is a reflection from optical glass components
in front of the CCDs.
The nearest and second nearest optical elements close to the CCDs
consist of the entrance window to the dewar and the filter;
their distances from the CCDs are 10.0mm and 34.5mm, respectively.
For both elements, the thickness and effective reflection index is 15.0mm and 1.460280, respectively.
The focal ratio Suprime-Cam is 1.86. Thus
light reflected on the first surface, the backside of the dewar window,
should be located  71.3 arcsec away from original position.
Light from the other surfaces appears  
144 , 212, 284 arcsec away from the original position.
\footnote{\citet{2002PASJ...54..833M} says the distance between the window and the filter
is 14.5 mm but this turns out to be modified to be 9.5 mm in the actual
implementation of Suprime-Cam. The translation of the flat glass (filter),
however, does not affect any optical performance.}
This model is consistent with the data and we use it to derive the mask radii.
Adopted mask radii for brighter stars are shown in Appendix \ref{tab:mask}.

We mask some stars with $R < 11$ by visual inspection to remove obvious artifacts.
Appendix \ref{tab:mask:GTO} and \ref{tab:mask:DLS} show the masks.

\section{Basic Weak Lensing Analysis}

To construct an initial weak lensing map we follow  the procedures described in \citet{2002ApJ...580L..97M,2007ApJ...669..714M} and \citet{2012ApJ...750..168K}. In this section we review the
selection of sources and the construction of the mass, noise, and  B-mode maps. In discussing
the B-mode map, we motivate the investigation (Section \ref{ImproveBmode}) of sources of systematic error in this map. 

\subsection {Sources}\label{sources}

\begin{figure}[h] 
   \begin{minipage}{0.6\hsize}
      \begin{center}
      \includegraphics[width=\hsize]
      {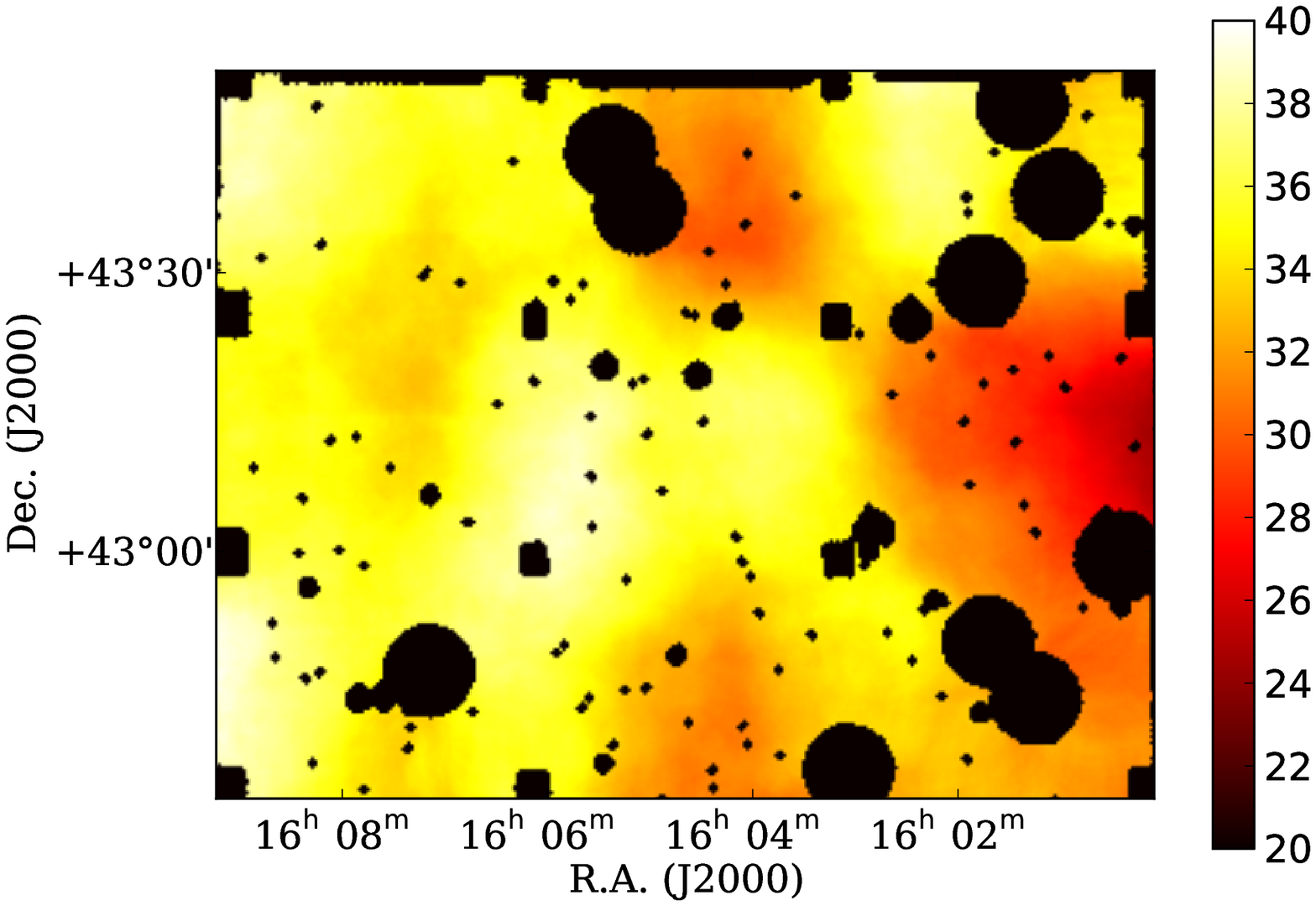} 
      \end{center}
   \end{minipage}
   \begin{minipage}{0.4\hsize}
      \begin{center}
      \includegraphics[bb= 300 70 700 500, clip, viewport= 0 100 350 600, width=\hsize]
      {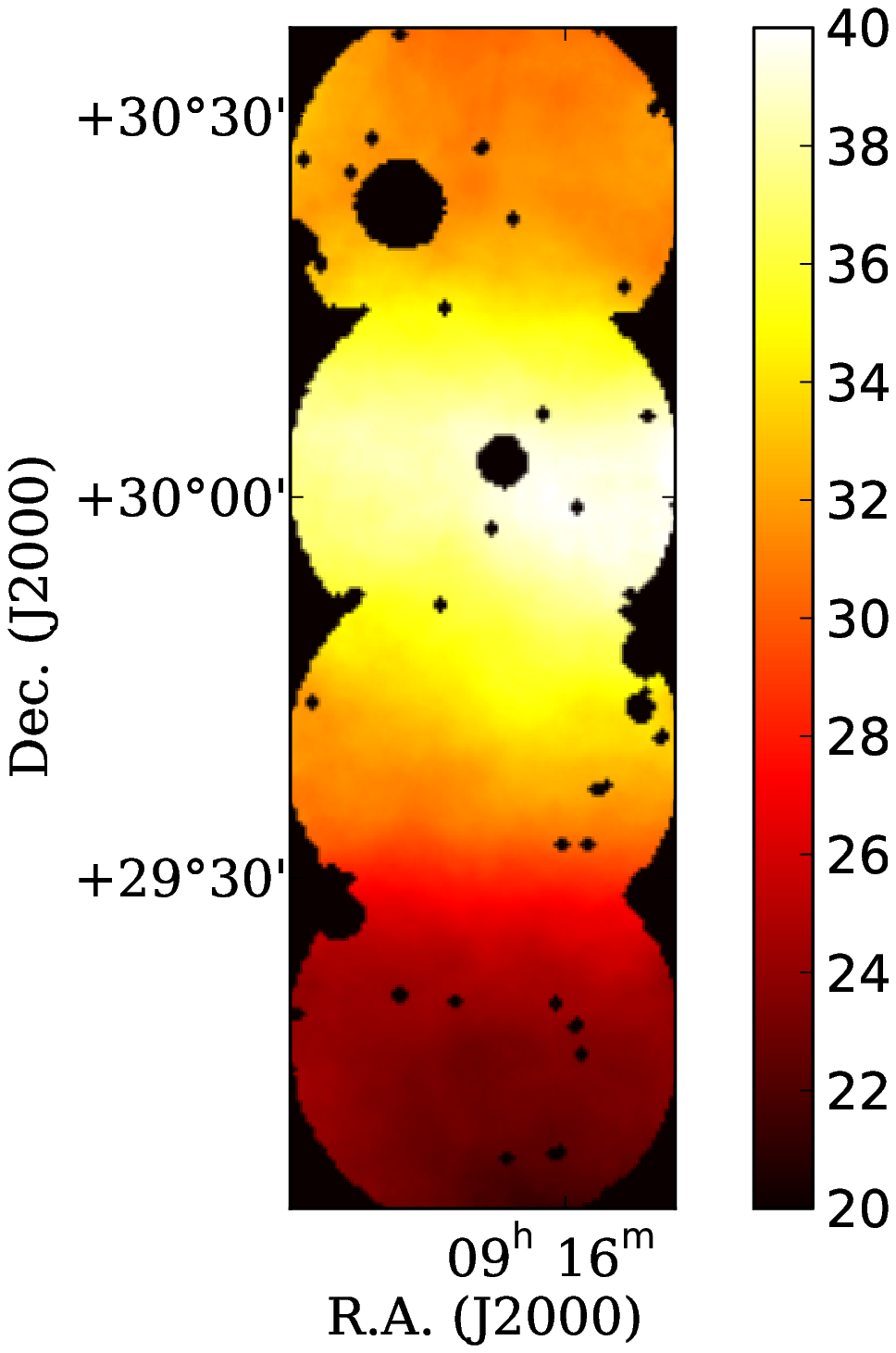}
      \end{center}
   \end{minipage}\\
   \caption
   {Surface number density of weak lensing sources. For The GTO field (left) the sources have $23 < R_c < 26$;
for the DLS field (right) the sources are in the range $23 < i^{\prime} < 26$.}
   \label{fig:DensityPlotForCMDSelectedGalaxies}
\end{figure}

We use $23<m<26$ galaxies with detection significance $\nu>10$ as source galaxies.
We eliminate  high ellipticity objects with  $|e|> 0.6$ because these objects are sometimes blended.
This magnitude range gives  an average magnitude of 24.5 for the sources in both the GTO and DLS fields. The average source density is $34.5{\rm~arcmin}^{-2}$ and $31.6{\rm~arcmin}^{-2}$ 
for the GTO and DLS fields, respectively.
Figure \ref{fig:DensityPlotForCMDSelectedGalaxies} shows  number density maps of sources galaxies for weak lensing analysis.

Because we do not have measured redshifts for the source  galaxies,
we use the fitting formula from \citet{2010A&A...516A..63S} for galaxy redshifts as a function of magnitude in the range
$23 < i_{814}<25$.
With this relation we estimate an average redshift of 1.2 for the DLS field where we have $i$-band data.
For the GTO field where we have only $R$-band date, there is no suitable redshift-magnitude relation.
We thus assume the same average redshift as the DLS field.
If the actual redshift is 1.0 rather than assumed 1.2, we introduce an $\sim10$\% error
in the estimate of the distance ratio.
We note that this error in mass estimation is only important for absolute calibration.

To derive the ellipticities of the sources, we use \emph{getshapes}, \emph{efit}, and \emph{ecorrect} in \emph{imcat}.
We evaluate the re-circularization correction
using stars in the field of view and we model the PSF distortion as a fifth order bi-polynomial as a function of position.
The $P_{\gamma}$ correction follows KSB+ \citep{1998ApJ...504..636H}
who adopt a galaxy size dependent correction.
The actual $\gamma$-field is discretized and noisy
because it can only be estimated at the positions of galaxies.
Note that we do not  weight the estimated shear value
according to galaxy's magnitude.
We apply a gaussian smoothing filter on the $\gamma$ field
following \citet{2011MNRAS.414.1851O}.
At first, we adopt a smoothing length, $\theta_{\rm G}=1.5^{\prime}$, following \citet{2012ApJ...750..168K}.

\subsection{Constructing the $\kappa$-S/N Map}

We can now construct the $\kappa$-S/N map to search for mass concentrations based
on the significance of the lensing signal.
We derive the $\kappa$ map directly from the $\gamma$ map of the sources in Section \ref{sources}.
We divide this map by the noise map to obtain the $\kappa$-S/N map.

We use Monte Carlo realizations to construct the noise maps.
We rotate the orientation of each source galaxy by a random angle.
The randomized catalog is then the basis for the noise map.
We repeat the Monte Carlo process 100 times,
and then calculate the rms at each grid point of the 
noise map to  estimate the noise at that point.
The $\kappa$-S/N map then reflects the noise from  intrinsic ellipticity noise
and from variation in the number of galaxies.

In the following sections (Section \ref{noisemap}, \ref{bmodemap}),
we  examine the systematic error in our weak lensing map
by looking at the  S/N distribution of the noise and B-mode maps.
We demonstrate that the noise map obeys the expected Gaussian S/N distribution;
however, the original B-mode map does not.

\subsection{Testing the Noise Map}\label{noisemap}

In the absence of signal, the S/N distribution for the noise map should match a standard normal distribution
which has no free parameters.
We test this expectation by constructing a ``Noise S/N map'' for each of the 100 Monte Carlo realizations.
Pixel values on each of these map  are strongly correlated spatially
because the maps are smoothed with the Gaussian kernel.
However a series of pixel values at a fixed position in the 100 ``Noise S/N maps'' 
should obey a normal distribution.
\begin{figure}[htbp] 
   \centering
   \includegraphics[width=\hsize, clip, viewport= 0 0 530 410,angle=-0]{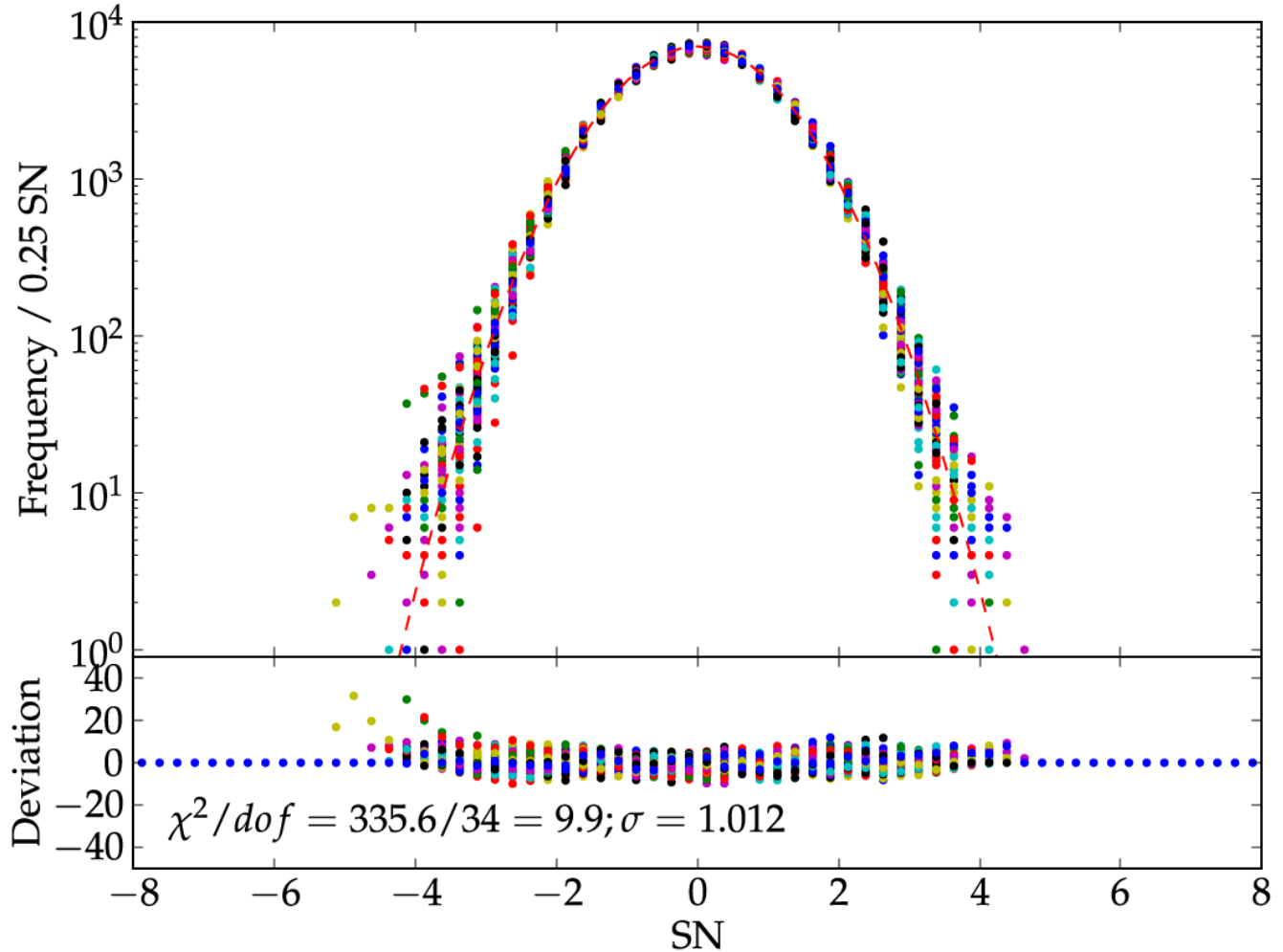}
   \caption{[Top] S/N distributions for 100 noise maps in the GTO field, which are
obtained by randomizing orientations of source galaxies. Different
points indicate different realizations. The dashed line shows the
expected normal distribution.
[Bottom] panel shows the deviation from the
normal distribution (Equation \ref{deviation}).
The sum of these deviations divided by the degree-of-freedom is a measure of the systematic error.
}
   \label{fig:noiseSNdistributionGTO1.5}
\end{figure}

\begin{figure}[htbp] 
   \centering
   \includegraphics[width=\hsize, clip, viewport= 0 0 530 410,angle=-0]{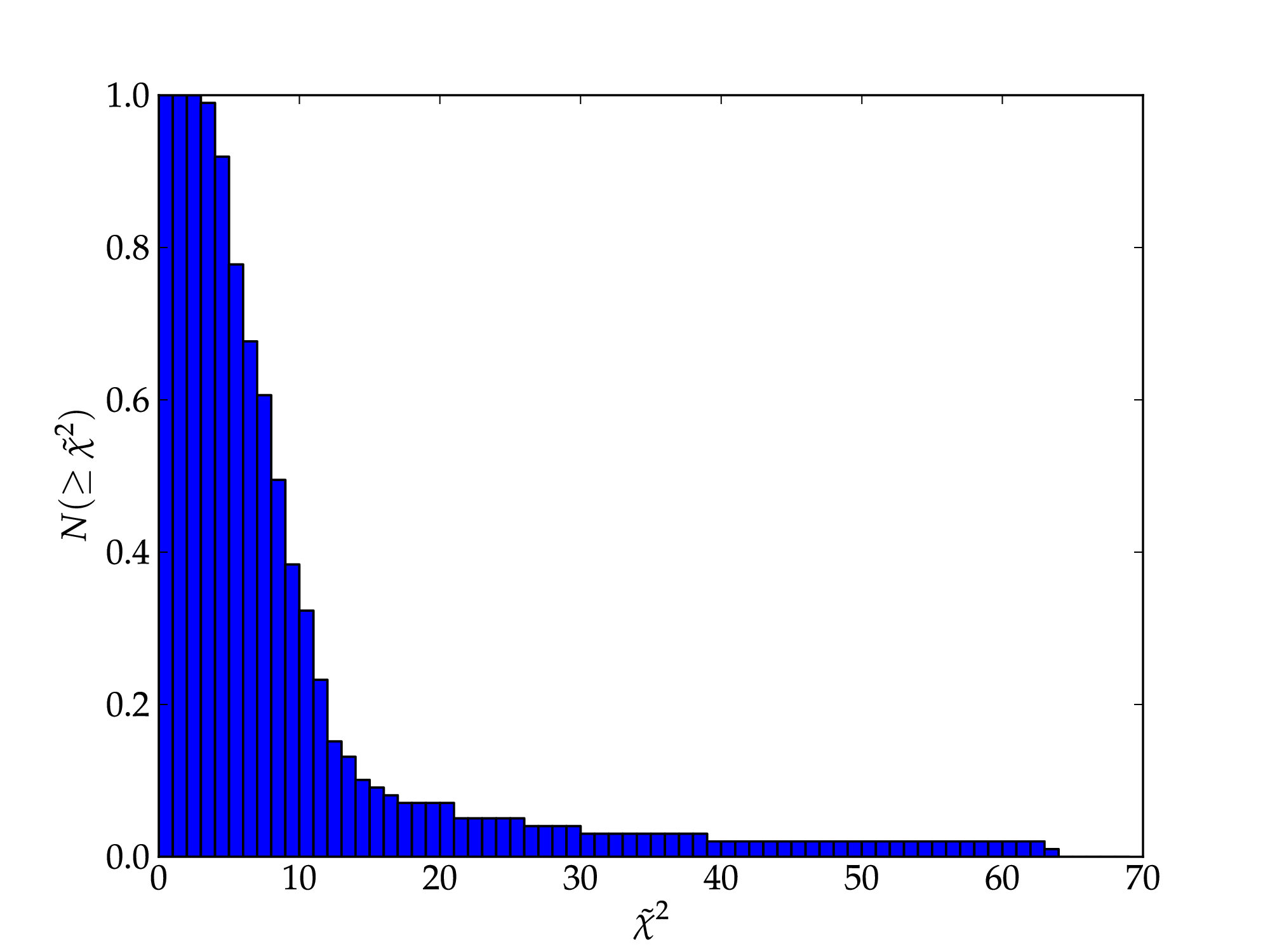}
   \caption{Normalized cumulative distribution of the merit function
$\tilde{\chi}^2$ (Equation \ref{reducedchi2}) from the 100 realizations of the noise
maps.}
   \label{fig:chidistGTO1.5}
\end{figure}

We show the S/N histogram for randomized realizations in Figure \ref{fig:noiseSNdistributionGTO1.5}.
The random realizations have some scatter that we use to evaluate the presence of systematic errors in the B-mode map.
To quantify the scatter, we define the ``Deviation'':
\begin{eqnarray}
	{\rm Deviation} = \frac{N_{i}^{\rm obs}-N_{i}^{\rm norm}}{\sqrt{N_{i}^{\rm norm}}}\label{deviation}
\end{eqnarray}
where $N_{i}=dN/d(S/N)_{i}$ is a number of pixels per $d(S/N)_{i}$ bin
in the  observed ({\it obs}) and in the standard normal distribution ({\it norm}).
The error is  the standard deviation.
We obtain the merit function as a kind of reduced chi-squared, $\tilde{\chi}^{2}$,
from a sum of the squared deviations dividing  by the the number degree of freedom (the number of bins):
\begin{eqnarray}
	\tilde{\chi}^{2} = \frac{1}{N-1}\sum_{i=0}^{N-1} \left(\rm {Deviation}\right)^{2}\label{reducedchi2}
\end{eqnarray}
Figure \ref{fig:chidistGTO1.5} shows the normalized cumulative
distribution of $\tilde{\chi}^{2}$ for all random realizations.
This wide range, $0\leq\tilde{\chi}^{2}\lesssim65$, gives the maximum allowable difference
between the noise and B-mode maps.

\subsection{Testing the B-mode Map}\label{bmodemap}
\begin{figure}[p] 
   \centering
   \includegraphics[width=\hsize]{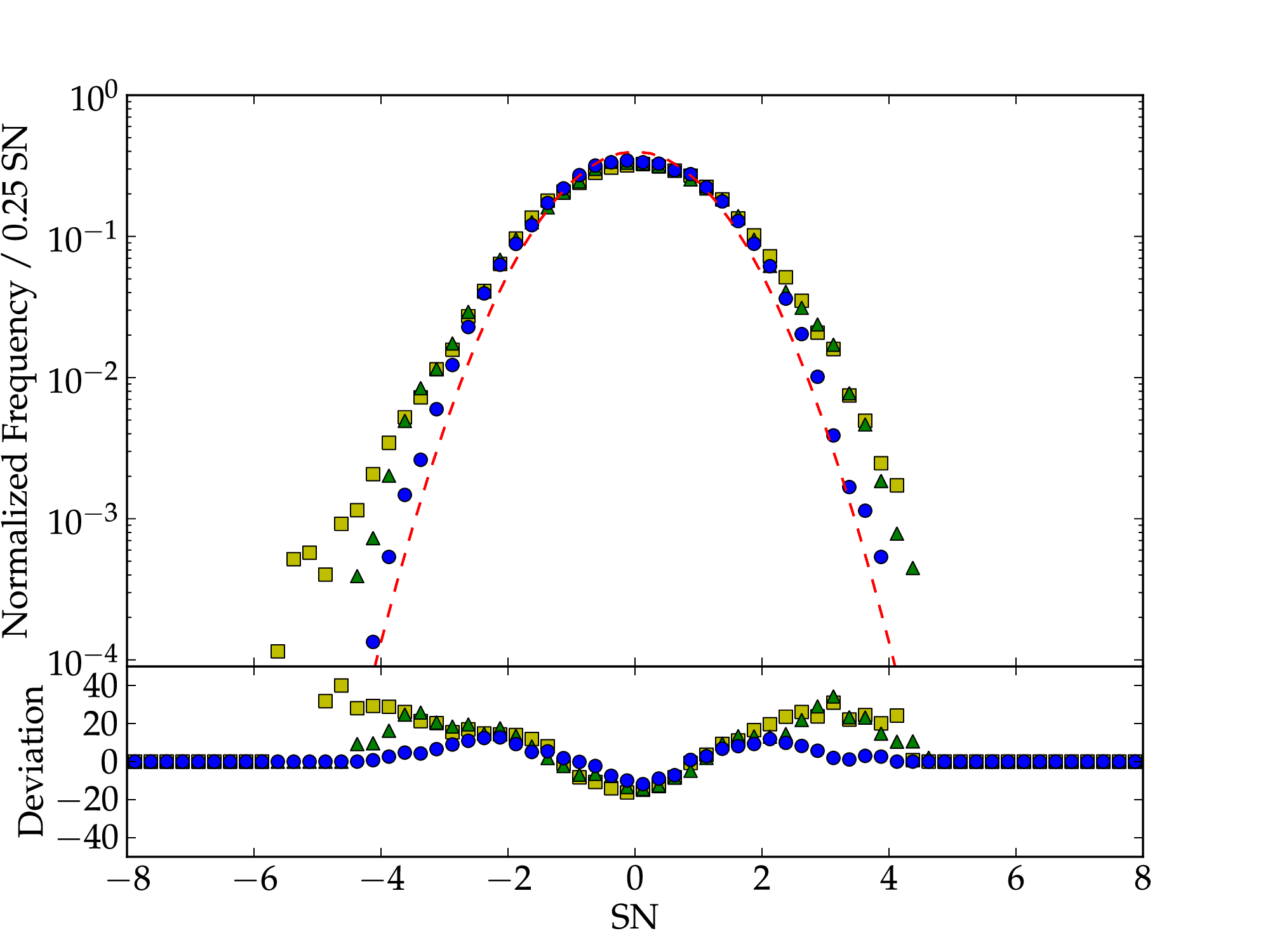} 
   \caption
   {[Top] Histogram of S/N values for B-mode maps derived from the new result ({\it circles})
   as well as original ({\it triangles}) and
   typical ({\it squares}) stacking. The dashed line shows the expected normal distribution if there is no systematic error.
   [Bottom] Same as \ref{fig:noiseSNdistributionGTO1.5} but for the three kinds of B-mode.
   }
   \label{fig:BmodeSNDistributionGTOOriginal}
\end{figure}
In this section, we quantify the quality of the B-mode map.
The standard B-mode map results from rotating each shear $|\gamma|\exp(2i\theta)$ by $\pi/4$,
i.e. $\gamma_{1}+i \gamma_{2}\rightarrow -\gamma_{2}+i \gamma_{1}$,
where $\theta$ is the position angle for each source.
Because weak lensing does not produce a B-mode,
we can evaluate 
the level of systematic error inherent in the mass reconstruction procedure by comparing the B-mode map with the noise map  \citep{2002A&A...389..729S}.

Figure \ref{fig:BmodeSNDistributionGTOOriginal} shows the pixel distribution  for the B-mode maps with  the
original, typical, and new stacking (Table \ref{tab:warprule}).
The S/N distribution has a large deviation relative to the standard normal distribution.
The $\tilde{\chi}^{2}$ = 257 (1204) 
for the original (typical) stacking.

A possible origin of these B-mode artifacts is  misalignment during the stacking procedure.
Assuming we are stacking circular gaussian objects, a
10\% positional inaccuracy generates $\sim 1$\%  artificial ellipticities.
The positional difference applies to individual galaxies and their neighbors.
The smoothing procedure then leads to detection of these $\sim1$\% ellipticities as a lensing signal.
Thus stars must be registered internally with an accuracy of 1\% level ($\sim$0.05 pixel). 

Next  we test procedures for reducing this kind of artificial signal in the B-mode map: 
(1) SkyFrame, (2) Wide FFTBox  (3) Masking the edge of the field,
(4) Higher order PSF anisotropy correction and (5) Smoothing cut off.

\section{Reducing the B-mode Signal}\label{ImproveBmode}
Systemic issues which might produce an artificial signal in the B-mode map include errors in the physical coordinate system,
excess large-scale power, boundary effects, and errors in the PSF anisotropy correction.

\paragraph{SkyFrame}
If the stacked image is bent with respect to the physical coordinate system,
the E-mode signal contaminates the B-mode. The tilt of the coordinates introduces a rotation of the sources.

The ``new stacking'' (Section \ref{newwarp}) eliminates this bending.
The ``new stacking''  image is aligned relative to the sky coordinate within $\sim1$ pixel;
the internal registration accuracy is $\sim0.05$ pixel.

\paragraph{Wide FFTbox}
There are two ways to perform the KSB analysis; (1)
convolve the shear field with the complex kernel
or (2) perform the convolution in the Fourier domain. Mathematically these procedures are identical.
However, in practice 
for operation in the Fourier domain,  boundary effects can be a problem.
To avoid boundary effects present in the original reduction \citep{2012ApJ...750..168K},
we embedded the shear signal for the E/B-mode maps
within a calculation box  twice the size of the field with zero padding outside the map area.

\paragraph{Masking the edge of the field}

At distances more than 15$^{\prime}$ from the optical axis
the field of view of Suprime-Cam  is vignetted \citep{2002PASJ...54..833M}.
The distortion is also large, $\gtrsim10^{\prime\prime}$, at the edge of the field.
We thus try masking the region outside 15$^{\prime}$.

\paragraph{Smoothing cut off}

To examine the origin of the remaining B-mode signal,
we examine the power spectrum of the B-mode.

\begin{figure}[htbp] 
   \centering
   \includegraphics[width=\hsize]{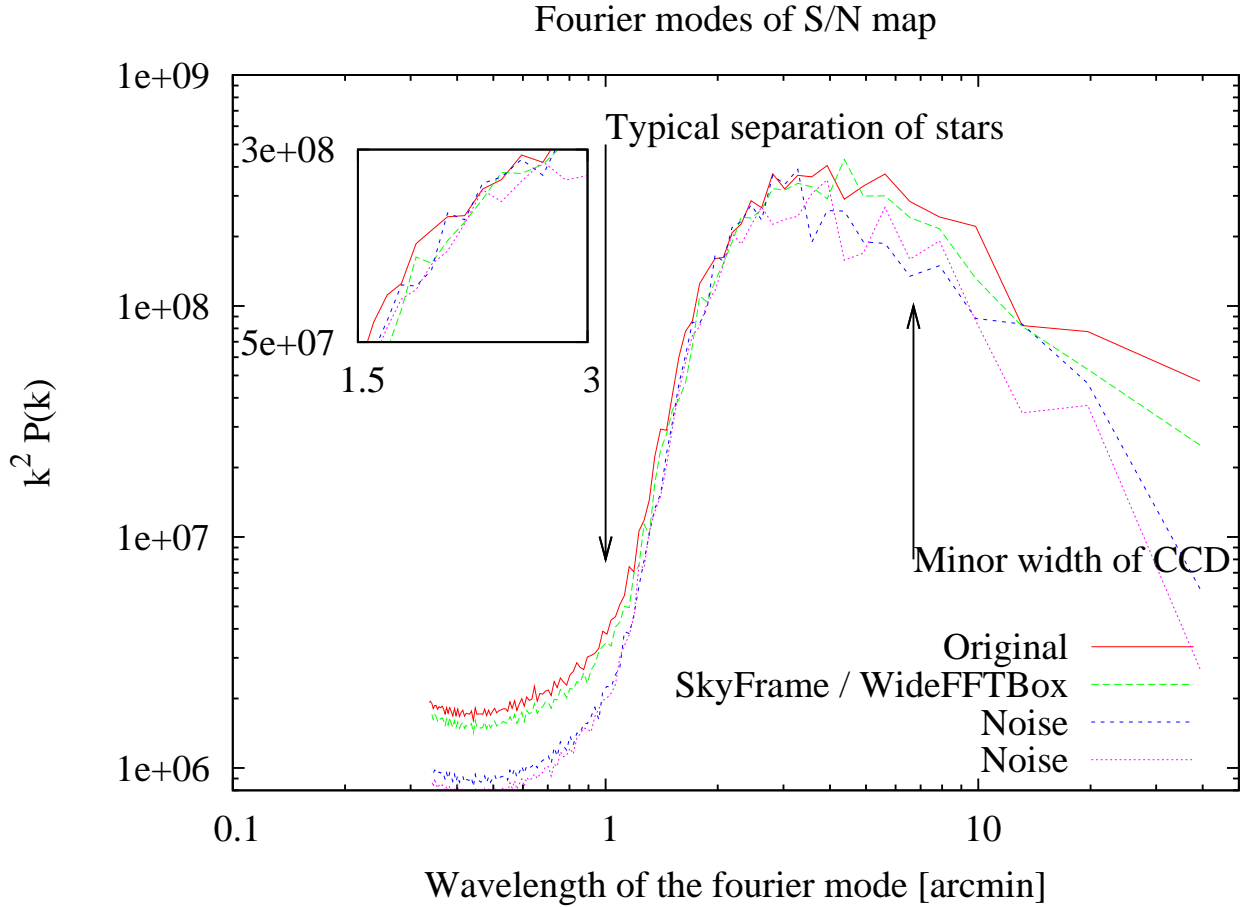} 
   \caption{Power spectrum of the noise maps (short dashed and dotted lines) and B-mode maps constructed with various procedures (all other curves as noted in the figure).
The power spectrum for the B-mode has an excess relative to the noise maps   on  smaller scale less than the typical separation of stars
   and on scales larger than 10 $^{\prime}$.
   For scales smaller than 1.5 $^{\prime}$,
   smoothing decreases the amplitude of the power spectrum.
   The inset shows the power spectrum in the range 1.5 to 3 $^{\prime}$.
   The SkyFrame procedure minimizes the excess in this range.
   A higher order PSF correction  has no impact.
   For scales of 1 -- 10 $^{\prime}$,
   there is no significant excess of in the B-mode power spectrum.
   For scales larger than 10$^{\prime}$, 
   all procedures fail to reduce the B-mode excess.
   Because  the typical scale of a cluster of galaxies is few arcminutes,
   these scale can safely be filtered out.
   }\label{fig:fftpower}
\end{figure}

There is a significant excess in the power spectrum of the B-mode map relative to the noise map on large scales ($\gtrsim10^{\prime}$). This excess is comparable with the power in the E-mode (Figure \ref{fig:fftpower}). The origin of this signal in the B-mode is unclear, but it should not be present.

To suppress this large scale systematic error in the B-mode map, we truncate the smoothing kernel at a scale of 10$^{\prime}$. Figure \ref{fig:BmodeSNDistributionGTOOriginal} shows the result. The truncation reduces the B-mode signal.

\paragraph{Higher Order PSF anisotropy correction}

To correct the PSF anisotropy induced by atmospheric disturbance or
optical aberration,  we applied a fifth order polynomial PSF model as a function of position.
If the order of the polynomial 
is not sufficient to remove the systematic variations in the PSF, a
B-mode is produced.
To investigate this issue,  we test a sixth-order polynomial.

\subsection{Summary of the Tests}\label{newprocedure}
\begin{table}[htdp]
\caption{Summary of the Tests}
\begin{center}
\begin{tabular}{l|c}
\hline
Tests	&	$\tilde{\chi}^{2}$\\
\hline
\hline
Typical stacking image	&	1204.7\\
Original result\footnote{\citet{2012ApJ...750..168K}}	&	257.7\\
\hline
SkyFame	&	90.7\\
SkyFrame+Wide FFTBox	&	81.9\\
SkyFrame+Wide FFTBox+6th&	84.3\\
SkyFrame+Wide FFTBox+Masking the edge	&	86.3\\
SkyFrame+Direct Convolution+Smoothing cut off	&	54.4\\
\hline
\end{tabular}
\end{center}
\footnotetext{Result values from the merit function defined as Equation \ref{reducedchi2} according to adopted tests.}
\label{tab:summarytests}
\end{table}%
Table \ref{tab:summarytests} shows the result of these procedures.
The ``SkyFrame'' registration  reduces the 
B-mode power on small scales and the ``Smoothing Cut-off'' reduces the power on large scales (Figure \ref{fig:fftpower}). These two procedures have the largest effect on reduction of systematics in the B-mode map.

The ``WideFFTBox'' also reduces the B-mode but not significantly. 
We thus use direct convolution instead of the FFT implementation.

``Masking the edge of the field'' and ``Higher order PSF anisotropy correction'' have little effect
for the GTO field. 
``Masking the edge of the field'' does not suppress systematics in the B-mode map. 
The same is true of the higher order corrections for PSF anisotropy.

Based on these results, we include 
\begin{itemize}
\item SkyFrame image
\item Smoothing Cut-off at 10 arcmin
\item Direct convolution instead of using Wide FFT Box
\end{itemize}
in the construction of the maps.
We show the result of this procedure in Figure \ref{fig:BmodeSNDistributionGTOOriginal}.
The pixel distribution of the B-mode S/N map indicates that
residual systematics are reduced to a reasonably low level if we adopt appropriate procedures.

To assess whether the deviation of the B-mode signal is consistent with random realizations,
we compare $\tilde{\chi}^{2}$ values with the distribution for each random realization.
The maximum $\tilde{\chi}^{2}$ is larger than what we derive from the B-mode signal for the GTO field
(Figure \ref{fig:chidistGTO1.5}),
$\tilde{\chi}^{2}=54.4$, much less than the original 257. Thus
we confirm that the our revised lensing map is substantially less affected by  systematic error.
Figure \ref{fig:MassMapGTO1.5} shows the resulting mass map with the ``new warp''
and Figure \ref{fig:BdistGTO1.5} shows the resulting S/N  distribution for the both mass and B-mode map.
\begin{figure}[p]
   \centering
   \includegraphics[angle=0,clip, viewport= 0 5 520 400, angle=-0]{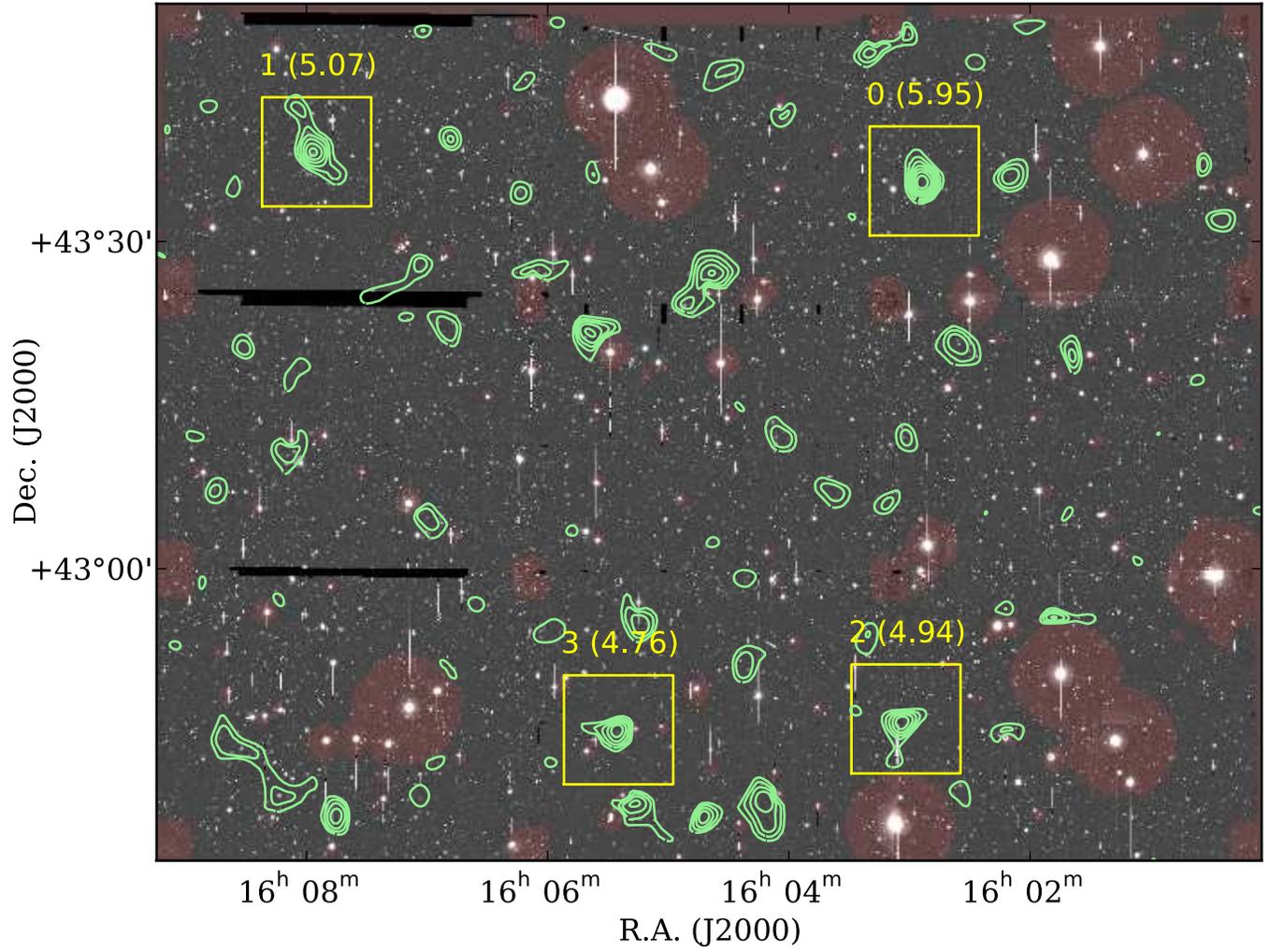}\label{fig:MassMapGTO1.5}
   \caption{
   Mass maps for the GTO field
   overlaid on Subaru images. Contours are drawn from $2\sigma$ threshold
   with an interval of $0.5\sigma$.  Boxes show identified peak positions
   along with peak number and its significance.
   Masked region for brighter stars are also overlaid.}\label{fig:MassMapGTO1.5}
\end{figure}
\begin{figure}[p]
   \centering
   \includegraphics[width=\hsize, clip, viewport= 0 0 530 410,angle=-0]{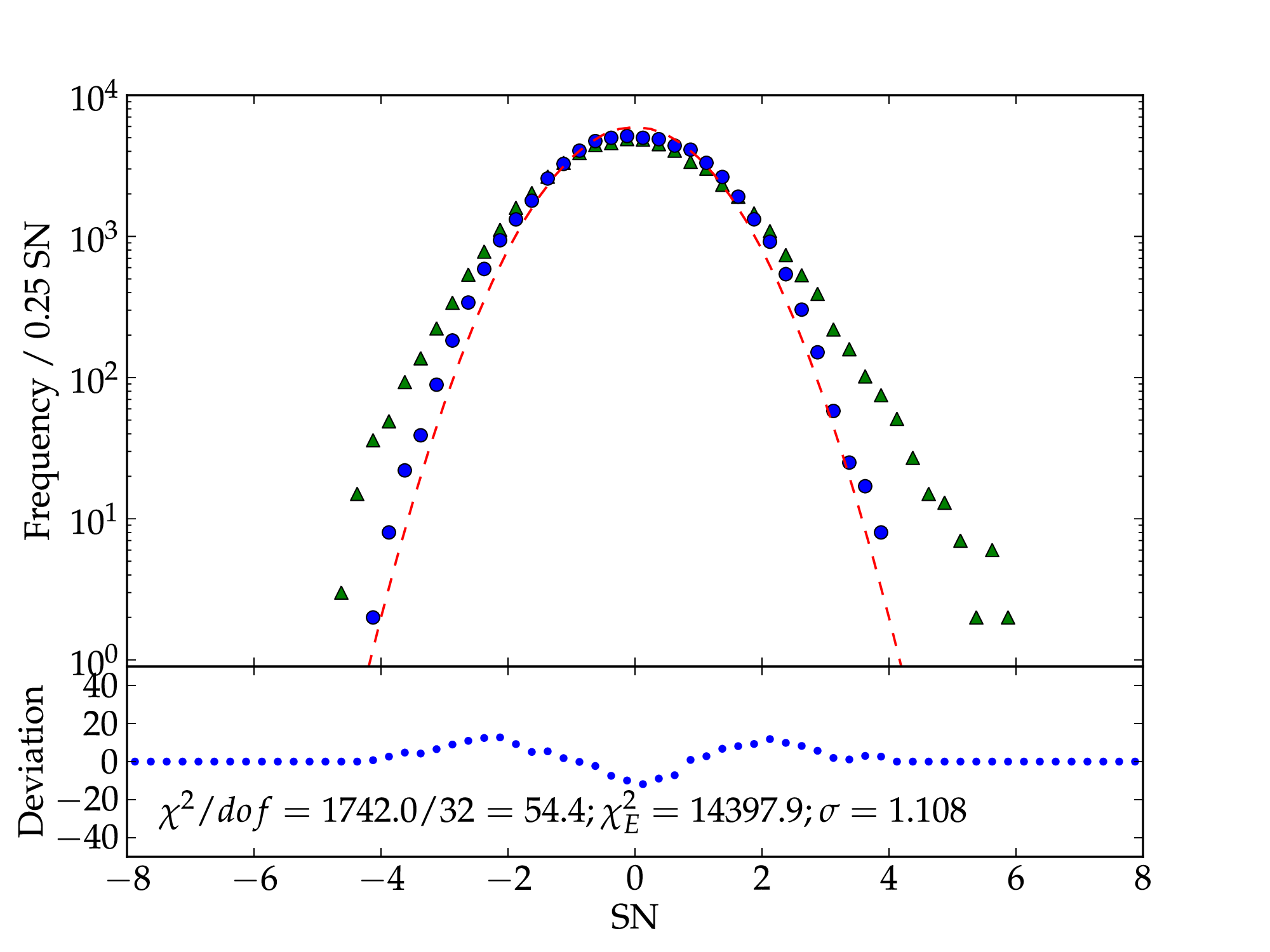}
   \caption{S/N distributions for E-mode ({\it triangles}) and B-mode ({\it
circles}) maps for the GTO fields. Lower panel shows the deviation from the
normal distribution for the B-mode map.}\label{fig:BdistGTO1.5}
\end{figure}

\subsection{The Impact of SkyFrame}

\begin{figure*}[p] 
   \centering
   \subfigure[GTO\_0]{\includegraphics[width=0.5\hsize]{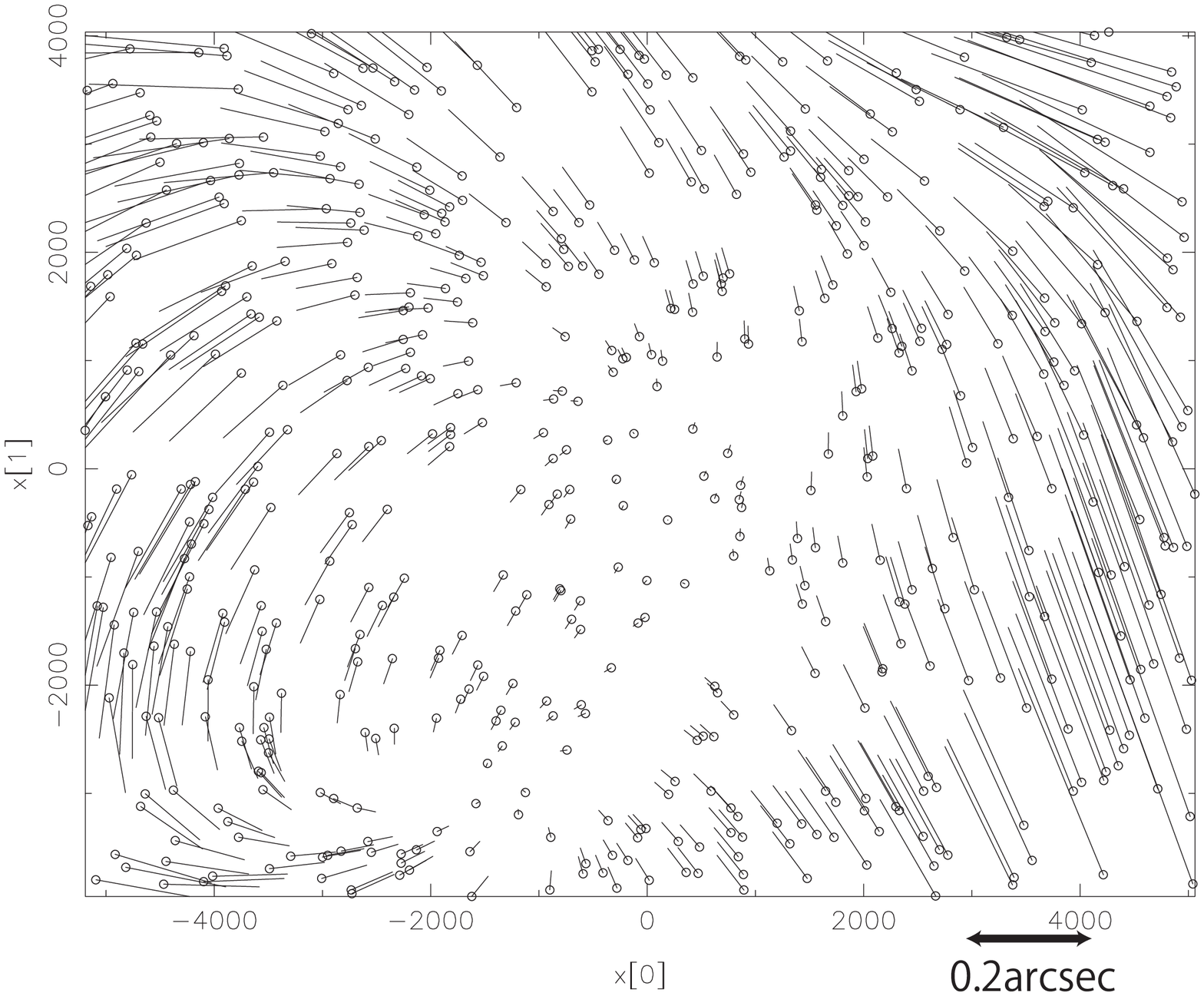}\label{fig:DisplacementSkyDetectorGTO_0}}
   \subfigure[GTO\_1]{\includegraphics[width=0.5\hsize]{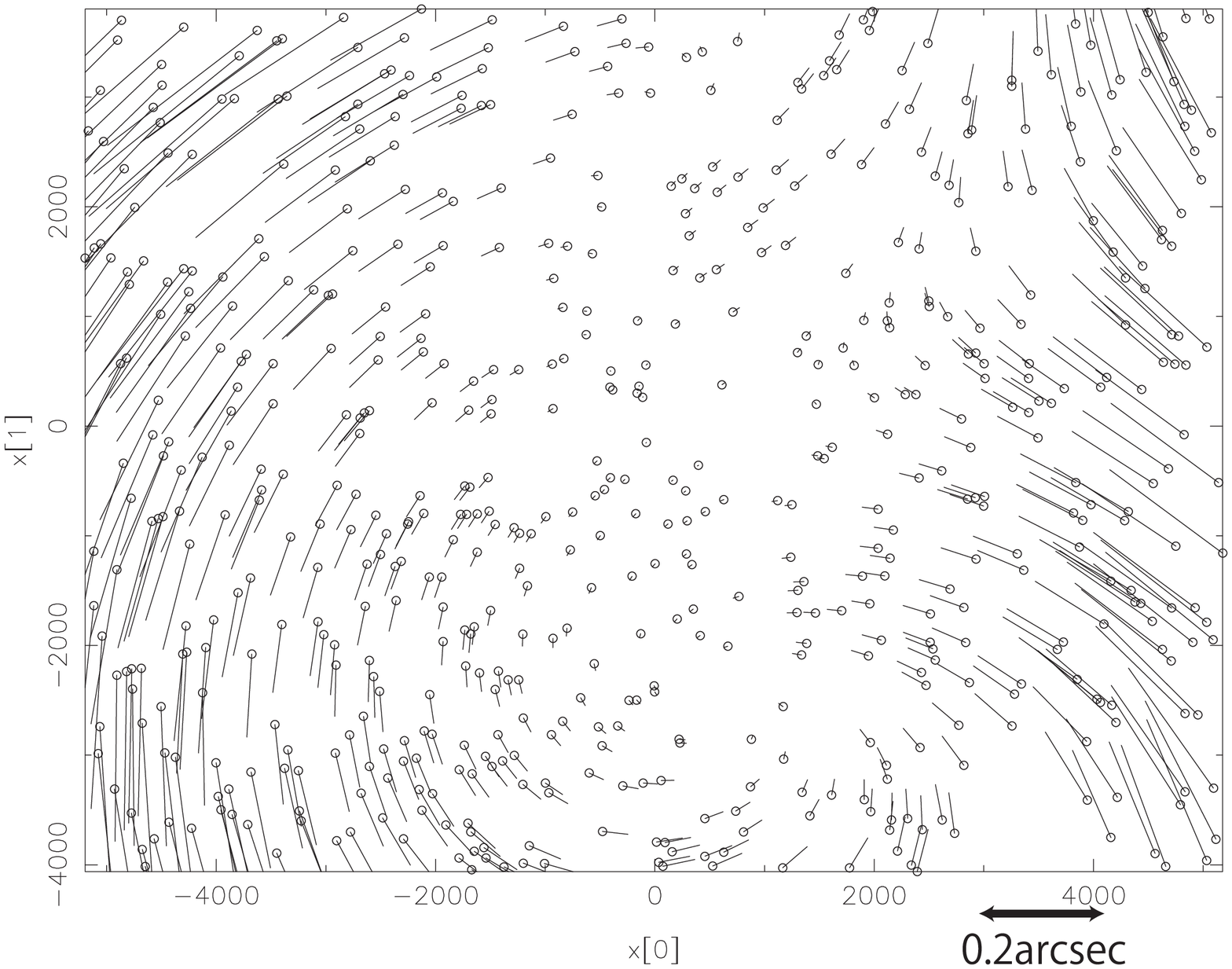}
   \label{fig:DisplacementSkyDetectorGTO_1}}
   \caption{
   Displacement vector map of stellar positional differences 
   between the undistorted detector coordinate (Section \ref{newwarp})
   and the stereographic projection of the celestial coordinate from the USNOA-2.0 catalog 
   at the tangent point for the GTO subfield.
   The box shows approximately $32'\times27'$ which is the field of view of Suprime-Cam.
   The length of the arrows indicate the positional difference in the detector coordinate multiplied by 1000.
   The  displacement is complex and includes  rotational components at the  level of the pixel scale.}
   \label{fig:DisplacementSkyDetector}
\end{figure*}
The most efficient procedure for  reducing the B-mode systematics is  ``SkyFrame''.
Figure \ref{fig:DisplacementSkyDetector} shows the
displacement vector of positional difference for stars
between the distortion free coordinate on the original warp
and the stereographic projection of the celestial coordinate from the USNOA2.0 catalog
at the tangent point.
Although the displacement is  small, a few pixels,
there is a coherent pattern.
Smoothing during construction of the map
can introduce coherent patterns that dominate  
the noise  from random ellipticities.
These patterns can introduce the small scale B-mode signal. 
It is also possible to generate false E-mode signal
from this coherent pattern.

There are two possible explanations for the residual pattern in the original warp:
(1) an anisotropic distortion resulting from  optical aberration, and/or
(2) a disturbance in the curvature of the wave front due to atmospheric effects.
The globally rotated component results from  a small rotation of the image rotator.
Higher order patterns vary from the field-to-field.

\citet{yoshino2007} evaluated the anisotropy in the distortion from the corrector lens in Suprime-Cam
with a positional accuracy of 0.4$^{\prime\prime}$.
They found that there is an $\sim$0.8$^{\prime\prime}$ difference
in the isotropic distortion pattern between 
the right  and  left edges of the field-of-view.
These distortions are comparable with the residual of the displacement 
from the distortion free coordinate on the original image to the projected celestial coordinate.

These patterns are different for different observed fields (Figure \ref{fig:DisplacementSkyDetectorGTO_0} and Figure \ref{fig:DisplacementSkyDetectorGTO_1}).
The difference between fields is the same as the difference of exposure-to-exposure
because all frames are simply registered to the first frame in the series of exposures.
These issues can be the source of the patterns we find.

A non-static factor like an atmospheric disturbance cannot explain the patterns.
\citet{2007ApJ...662..744D} show that longer exposures 
yield progressively lower ellipticities of the PSF (as $1/\sqrt{t_{\rm exp}}$ )
suggesting that the coherent differences shown in Figure \ref{fig:DisplacementSkyDetectorGTO_0} should not be caused from  atmospheric disturbance.

\section {Defining the Weak Lensing Peaks}

At least two issues are important in defining the weak lensing peaks in the $\kappa$-S/N map. The smoothing length is important because it reduces the shape noise. Furthermore, the smoothing in part determines the sensitivity to cluster detection as a function of mass and redshift. The threshold for peak finding is also important. The fraction of false positives is sensitive to the threshold. Here we discuss these two issues. 

\subsection{Smoothing Length}\label{smoothinglength}

In the previous sections, we  constructed weak lensing mass maps with a fixed smoothing length of 1.5 $^{\prime}$.
A smaller smoothing length  introduces more noise in the map resulting from shape noise.
A larger smoothing length (larger than a few arcminutes)  smooths out the cluster signal;
the angular diameter of typical clusters for redshifts $0.2<z<0.7$, the range of sensitivity of the
weak lensing maps, is several arcminutes.

Previous studies adopt a range of smoothing lengths.
\citep{2001ApJ...557L..89W,2002ApJ...580L..97M,2005A&A...442...43H,2006ApJ...643..128W,2007A&A...462..875S,2007A&A...462..459G,2007ApJ...669..714M,2007A&A...462..473M,2008MNRAS.385..695B,2007A&A...470..821D}.
To examine  the effect of the smoothing length,
we construct E/B-mode  maps and random maps with three different smoothing lengths:
$\theta_{\rm G}$ = 0.5$^{\prime}$, 1.0$^{\prime}$, and 1.5 $^{\prime}$.

\begin{figure*}[p] 
   \centering
   \includegraphics[width=\hsize]{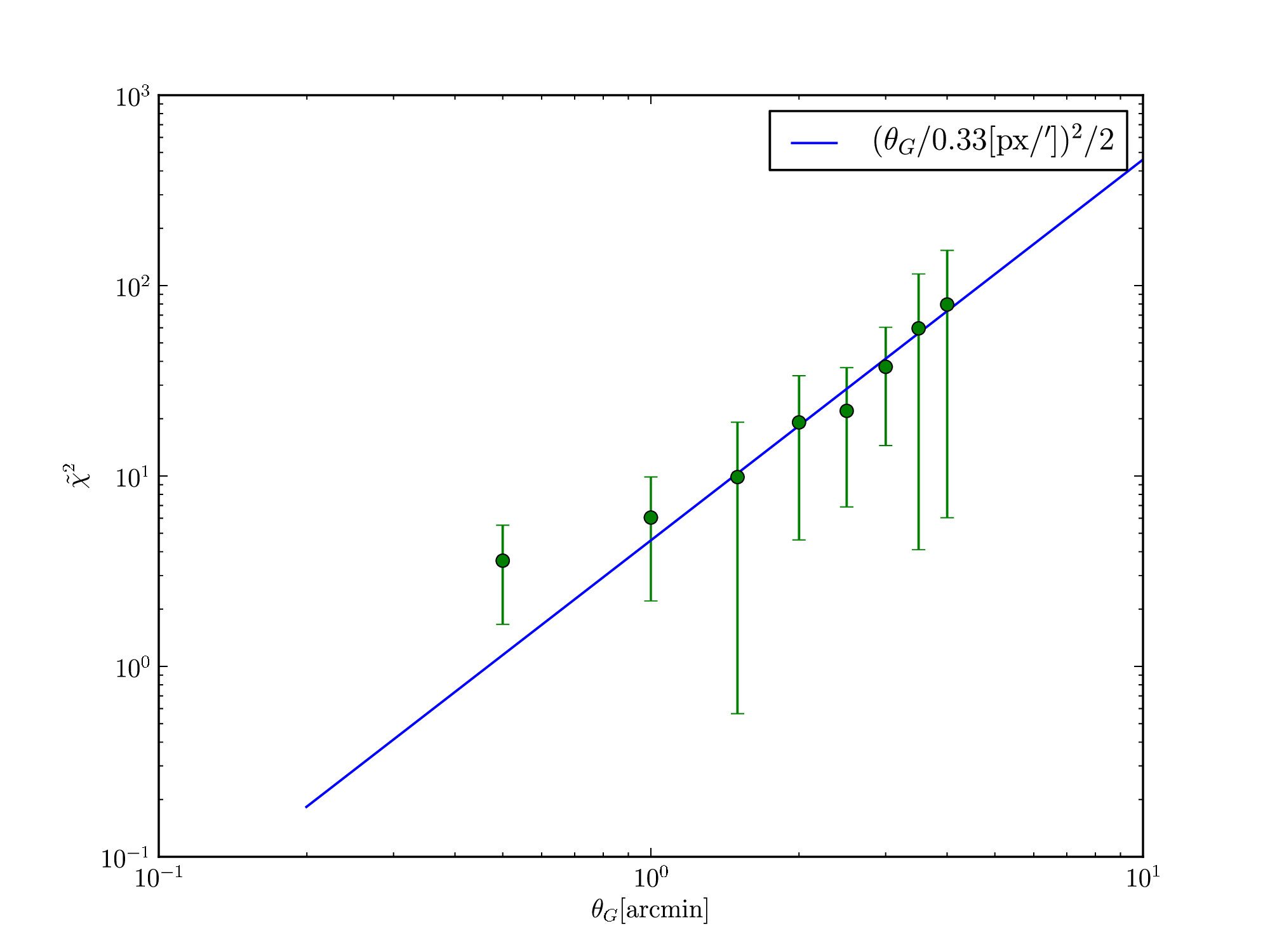} 
   \caption{Relation between the merit function $\tilde{\chi^{2}}$, and the smoothing scale $\theta_{\rm G}$.
   Points with errors denote mean values of $\tilde{\chi^{2}}$ and its standard deviation.
   The solid line shows $(\theta_{\rm G}/0.33 [{\rm px}/'])^{2}/2$,
   the square of the sigma of the Gaussian smoothing kernel in pixel units.}
   \label{fig:chisqtg}
\end{figure*}

Figure \ref{fig:chisqtg} shows the relation between $\tilde{\chi}^{2}$ and $\theta_{\rm G}$.
The $\tilde{\chi}^{2}$ of the E-mode increases with the smoothing length. The $\tilde{\chi}^{2}$ 
of the B-mode also increases.

The averaged $\tilde{\chi}^{2}$ value among 100 random realizations is $\tilde{\chi}^{2} > 1$ and it
increases with 
the square root of the smoothing length $\theta_{\rm G}$.
This result occurs because the smoothing reduces  the effective number of independent pixels.
For small $\theta_{\rm G}$ the behavior of $\tilde{\chi}^{2}$
comes from the pixelization caused
by dividing the smoothing kernel into coarse pixels.
The shortest smoothing length, $\theta_{\rm G}=0.5'$, corresponds to only $\approx1.5$ pixels.
The coarse sampling cannot reproduce the gaussian smoothing kernel well; thus
the amplitude of the scaling relation may be changed and increased.
This kind of pixelization effect is hard to describe analytically.
Because the range, $0'.5<\theta_{\rm G}<1'.5$, might be affected by this pixelization effect,
we do not correct for the number of effective pixels. We use the scale dependent $\tilde{\chi}^{2}(\theta_{\rm G})$
as a measure of the $\tilde{\chi}^{2}$.

We estimate the optimal smoothing length for detecting clusters as a function of mass and redshift.
Following \citet{2004MNRAS.350..893H},
we  analytically evaluate the expected S/N  for
weak lensing halos with universal NFW profiles.
Because more distant clusters have smaller angular diameter,
the S/N is redshift dependent.
Furthermore, the typical scale radius of an NFW cluster depends on its mass;
thus the S/N is also mass dependent.
We compute the S/N as a function of smoothing scale $\theta_{\rm G}$.
The S/N for less massive clusters
($M_{14}\equiv10^{14}h^{-1}M_{\odot}$; $M_{14}=3.0$ for $z=0.5$, $M_{14}=1.0$ for $z=0.2$)
reaches a maximum at $\theta_{\rm G}\approx1.0'$;
more massive clusters are detected at greater S/N with a smoothing length larger than $\theta_{\rm G}\approx1.0'$.
However, difference between 1.0$^{\prime}$ and 1.5 $^{\prime}$ is small.
At a fixed cluster mass, smaller smoothing lengths yield higher S/N at larger redshift.
Because these effects are small, we choose $\theta_{G}=1.5'$ .

\subsection{Peak Definitions and Thresholds}
\begin{figure}[htbp] 
   \centering
   \includegraphics[width=\hsize, clip, viewport= 0 10 530 410,angle=-0]{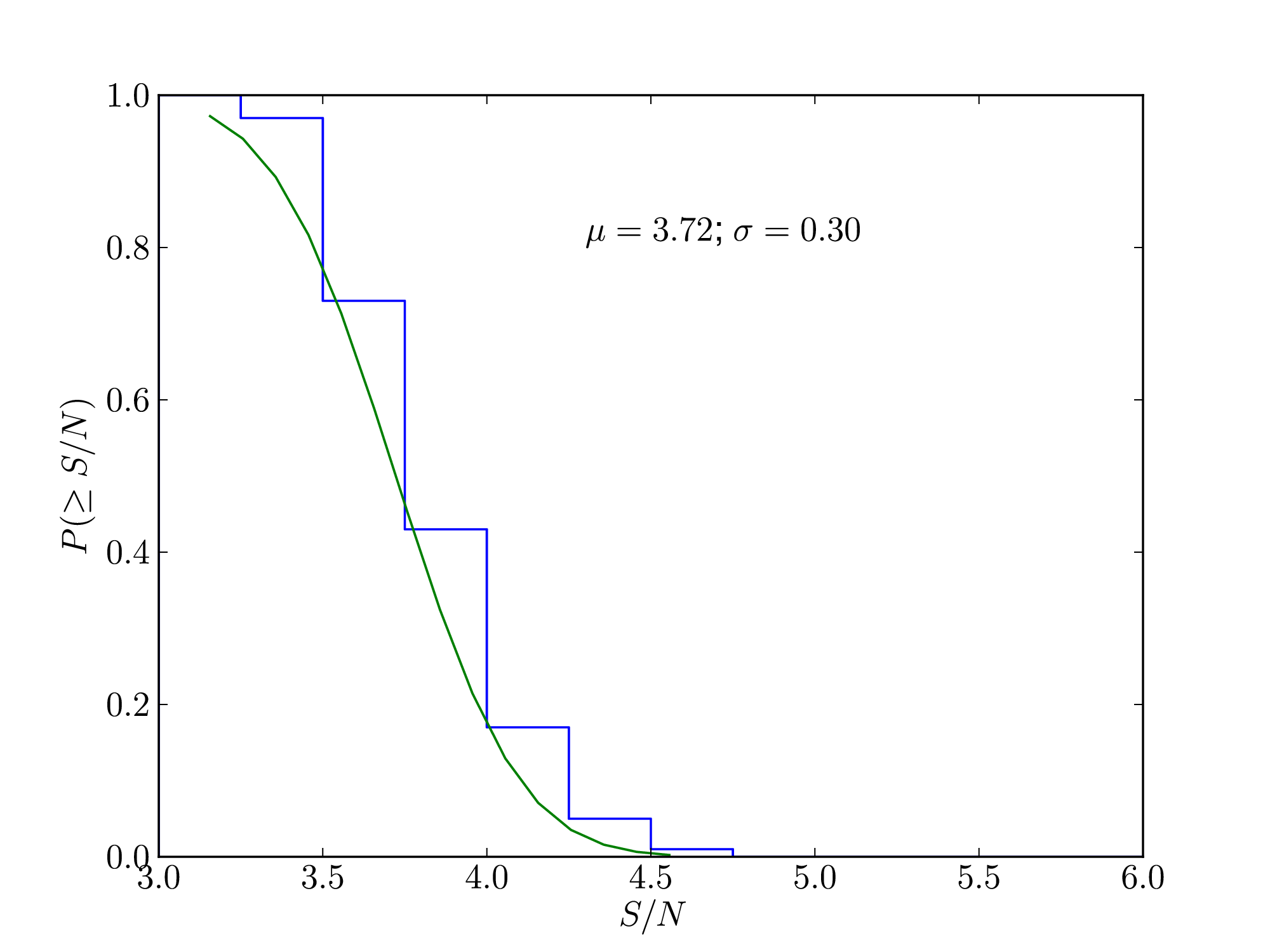}
   \caption{Cumulative distribution of the highest peak S/N values for 100
realizations of noise maps for the GTO field. The solid line shows the
expectation from the normal distribution.}
   \label{fig:randpeakheightsGTO1.5}
\end{figure}

We use SExtractor to locate peaks in the E-mode map; we require
at least 3 connected pixels ($>2\sigma$) to define a peak.
Selecting the appropriate threshold for a clean peak catalog is s subtle issue. 
Figure \ref{fig:randpeakheightsGTO1.5} shows the cumulative histogram of the highest peaks
found in each noise map as a function of S/N.
This figure shows the likelihood that  significant peaks can be found accidentally. 
Statistics describing the highest peak for the 100 realizations
are shown in the figure and summarized in Table \ref{tab:varyingthg}.

We compare the cumulative  peak counts with the expected standard normal distribution (green line).
Figure \ref{fig:randpeakheightsGTO1.5} shows that the highest peak statistics of the noise maps are not well described by the theoretical curve. In other words,
the highest peak statistics do not obey the ideal random Gaussian field. Thus we must
derive these statistical quantities from  Monte-Carlo simulations.

This discrepancy between the noise maps and the ideal case is more significant for smaller smoothing lengths.
Based on these tests, we select different thresholds for each  smoothing length.
We consider two kinds of threshold: the $3\sigma$ threshold and the 99\% threshold.
The $3\sigma$ threshold is derived by assuming the standard normal distribution for the highest peaks;
the 99\% threshold is derived from ``the most extreme peaks'' among the highest peaks in the entire set of 100 noise map.
Because it is based on the real data, the 99\% threshold is much more reliable than the $3\sigma$ threshold and we use it throughout.

We summarize the $3\sigma$ and 99\% thresholds in Table \ref{tab:varyingthg}.
As expected these thresholds differ
for  smaller  smoothing lengths, but they are similar
for  $\theta_{G}=1.5$.

The numbers of detected peaks for each threshold
are listed as $N_{3\sigma}, N_{99\%}$  in Table \ref{tab:varyingthg}.
\begin{table*}[htdp]
\caption{S/N and Peak Statistics as a Function of the Smoothing Length $\theta_{\rm G}$}
\begin{center}
\begin{tabular}{c|ccc|cccc|cc}
	&	\multicolumn{3}{c|}{S/N values}	&	\multicolumn{4}{c|}{Max Peak}	&	\multicolumn{2}{c}{Detected Peak}\\
\hline
$\theta_{\rm G}$	&	$\tilde{\chi}^{2}_{B}$	&	$\tilde{\chi}^{2}_{E}$		&	$\sigma$	&	$\mu$	&	$\sigma$	&	$(S/N)_{3\sigma}$	&	$(S/N)_{99\%}$	&	$N_{\rm 3\sigma}$	&	$N_{\rm 99\%}$\\
\hline
\hline
0.5	&	1.94	&	29.9	&	1.01	&	4.07	&	0.25	&	4.82	&	5.05	&	2	&	0\\
1.0	&	15.1	&	175	&	1.05	&	3.83	&	0.22	&	4.48	&	4.64	&	4	&	4\\
1.5	&	54.4	&	545	&	1.11	&	3.72	&	0.30	&	4.62	&	4.57	&	4	&	4\\
\hline
\end{tabular}
\end{center}
\footnotetext{``S/N values'' shows  statistical properties for the S/N histogram, the merit functions for E/B-mode, and the standard deviation of pixel values. ``Max Peak'' shows  statistical properties for the highest peak value $(S/N)_{3\sigma}$ and the threshold given the most extreme peak among the highest peaks in the entire set of 100 noise map ($(S/N)_{\rm 99\%}$). ``Detected Peak'' show the number of detected peaks corresponding to the adopted threshold.}
\label{tab:varyingthg}
\end{table*}%

Next we consider  the peaks in the E-mode map.
With the smallest smoothing, there are no peaks. 
The shape noise of intrinsic ellipticities raises the threshold.
The situation is different for $\theta_{G}=1.0$ and 1.5
because the larger smoothing involves more sources.
The lensing signal becomes more significant for larger smoothing lengths.

As a quantitative representation of the significance of the weak lensing signal,
we evaluate $\tilde{\chi}^{2}_{E}/\tilde{\chi}^{2}_{B}$.
For the B-mode, $\tilde{\chi}^{2}$ should be small because there should be no signal;
for the E-mode, $\tilde{\chi}^{2}$ should be larger reflecting the lensing signal.

The actual values of $\tilde{\chi}^{2}_{E}/\tilde{\chi}^{2}_{B}$ for $\theta_{G}=1.0$ and 1.5 are 11.6 and 10.0 for the GTO field.
These similar values imply
there is no strong preference for  smoothing lengths between 1.0 and 1.5 in our data.

\section{The DLS kappa Map}

The procedures of Sections \ref{newprocedure} generally apply to both the GTO and DLS field maps. Here we discuss
the few issues that are particular to the DLS map.

\begin{figure}[p]
   \centering
   \includegraphics[width=\hsize, clip, viewport= 0 0 530 410,angle=-0]{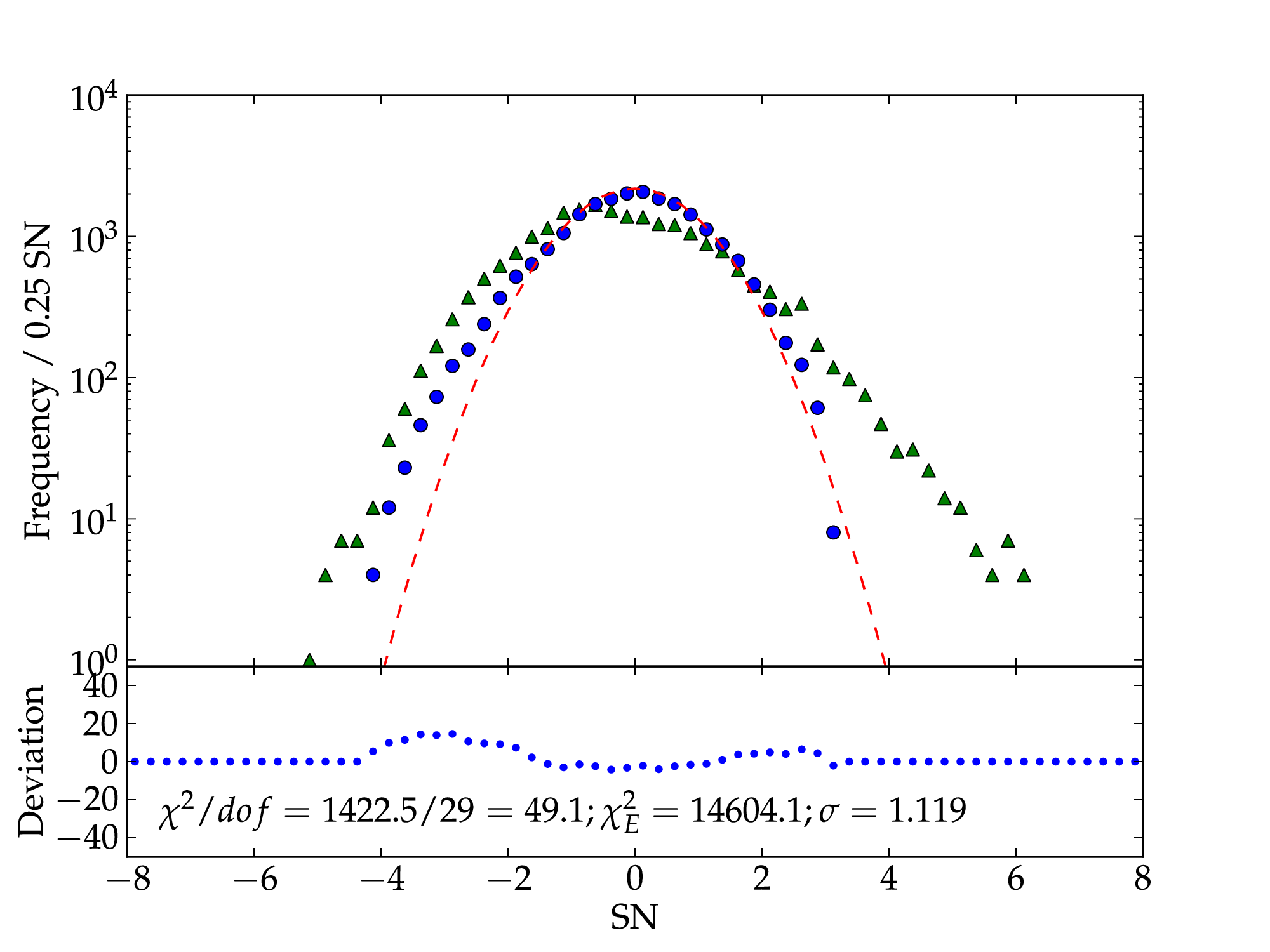}
   \caption{Same as Figure \ref{fig:BdistGTO1.5} but for the DLS field.}\label{fig:BdistDLS1.5}
\end{figure}
For the DLS map, the  $\tilde{\chi}^{2}$ for the original procedure is 134; for the new one it is
49.1 (Figure \ref{fig:BdistDLS1.5}). The new value although reduced from the original one is not within the 
range for the random realizations (32 is the maximum value).
The S/N distribution for the B-mode has a slight excess on the negative side.

We use $i'$ band imaging in the DLS field rather than the $R_{C}$ band.
We cannot determine the PSF anisotropy near the edge of the field because the star/galaxy separation is poor.
Thus we trim objects located at the outer edge ($>15^{\prime}$ of the center of field). In contrast we do not trim the edges of the GTO field.

\begin{figure}[p]
   \centering
   \includegraphics[angle=0, clip, viewport= 0 5 520 400, angle=-0]{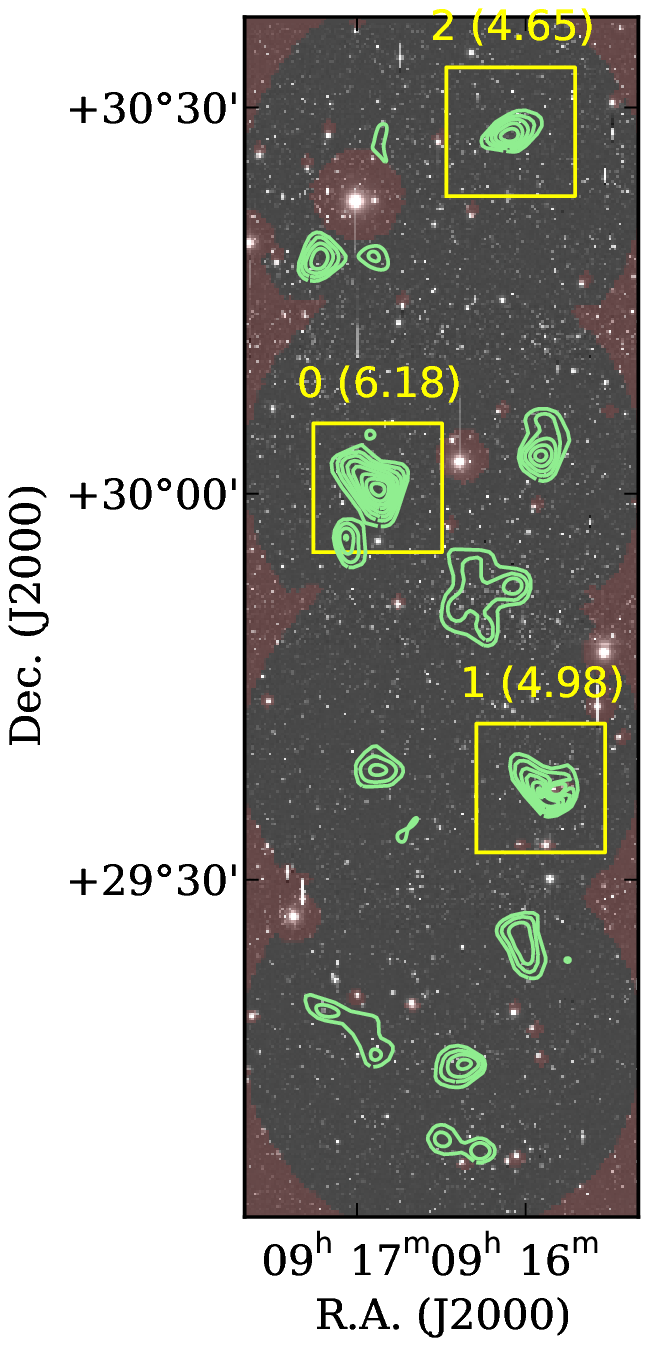}
   \caption{
   Same as Figure \ref{fig:MassMapGTO1.5} but for the DLS field.}\label{fig:MassMapDLS1.5}
\end{figure}

Even after this trimming, there is still
a small excess in the S/N distribution for the B-mode on the negative side.
However, the plus side seems to be well matched to the standard normal distribution.
If we mask the three prominent negative excesses,
the B-mode S/N distribution is in good agreement with the standard normal distribution.
Thus the global  correction may be sufficient and
the residual local excess may be related to the most prominent peaks in the E-mode map.
Figure \ref{fig:MassMapDLS1.5} shows the $\kappa$-S/N map for the DLS field.
Here we present the resulting $\kappa$-S/N maps for the DLS field (Figure \ref{fig:MassMapDLS1.5})

\section{Distribution of Peaks in Noise and B-Mode Maps}

It is useful to have an analytical expression for the distribution of the highest peaks in order to test both the noise and B-mode maps. These expressions can then be used to estimate the fraction of false peaks in the
$\kappa$-S/N maps.

We derive a probability density function that enables us to evaluate
the probability of finding the highest peak in the noise map as a spurious peak in the $\kappa$-S/N map. We evaluate this probability as a function of the S/N.
The probability density function that a random value $x$ obeying the normal distribution becomes the maximum value among
$n$  samples is:
\begin{eqnarray}
	P_{n}(x)dx = \frac{n}{\sqrt{2\pi}} \exp\left( -\frac{x^{2}}{2}\right)\left(\frac{1}{2}{\rm erfc}\left(-\frac{x}{\sqrt{2}}\right)\right)^{n-1} dx
\end{eqnarray}
where ${\rm erfc}(x)$ is the complementary error function: ${\rm erfc}(x) = 2/\sqrt{\pi} \int^{\infty}_{x}e^{-t^{2}}dt$.
The first part of the product gives the probability of obtaining a random value of $x$,
and the latter part describes the probability that the remaining $n-1$ random values are $< x$.

However, our map creation procedure is more complex because it applies gaussian smoothing which correlates pixel values spatially.
We have to check the effect of the spatial correlation prior to evaluating the probability.
We apply the smoothing function on the $\kappa$ map in the form $\exp\left(-(\theta/\theta_{\rm G})^{2}\right)$ where
$\theta_{\rm G}/\sqrt{2}$ is the standard deviation of the smoothing kernel.
Using the standard deviation as a measure of the correlation length,
we reduce the total number of pixels in the mass map  by a factor of $(\theta_{\rm G}/\sqrt{2})^{2}$.
For example, in the GTO field where the total number of pixels is $300\times234$ with a 0.33$^{\prime}$ / pixel scale,
the effective number of pixels as a result of the smoothing becomes
$300\times234 \times (0.33 \times \sqrt{2}/\theta_{\rm G})^{2}$.
After adopting $\theta_{\rm G}=1.5^{\prime}$, the effective number of pixels is 6,795.

\begin{figure}[htbp] 
   \centering
   \includegraphics[width=6in]{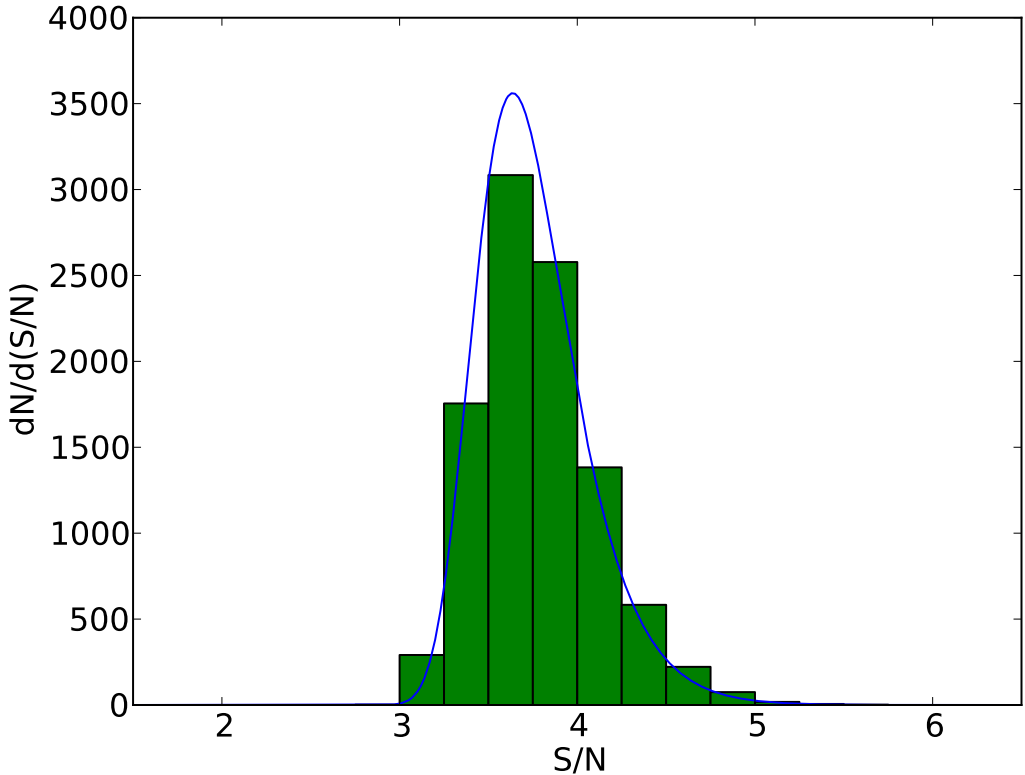} 
   \caption{S/N value distribution for maximum peaks among 10,000 noise map realizations for a
   field covering 2.12 deg$^2$ ({\it histogram}) and an analytical model ({\it curve}). The effective pixel number
   is the number of actual pixels divided by the smoothing length in pixel units.}
   \label{fig:SNanalytical}
\end{figure}

Figure \ref{fig:SNanalytical} shows  both the analytic curve  and the Monte Carlo results based on 10000 random realizations of the noise map.
The analytic expression is an excellent representation of the Monte Carlo results.

In summary, Figure \ref{fig:SNanalytical} shows that the smoothing
procedure only  reduces the number of independent pixels.
Using this effective number in the probability function, 
we can evaluate the confidence level corresponding to a given  S/N threshold. For example,
a threshold of S/N = 4.66 corresponds to a 99.7\% confidence level for the 2.12 deg$^{2}$ B-mode free massmap based on 1.5$^{\prime}$  smoothing.
Similarly, S/N = 4.50 and 5.80 are the same 99.7\% level confidence levels for a 1 and 1000 deg$^{2}$
degree survey, respectively.

Using this same formalism, we can derive an expression for the fraction of spurious peaks in the $\kappa$-S/N map above a given threshold.
The probability of obtaining the $(m+1)$-th value in the descending set of ordered random values as a function of S/N is:
\begin{eqnarray}
	P_{n}^{(m)}(x)dx &=& \frac{(n-m) {}_{n}C_{m}}{\sqrt{2\pi}} \exp\left( -\frac{x^{2}}{2}\right)
	\left(\frac{1}{2}{\rm erfc}\left(-\frac{x}{\sqrt{2}}\right)\right)^{n-m-1}
	\left(1-\frac{1}{2}{\rm erfc}\left(-\frac{x}{\sqrt{2}}\right)\right)^{m} dx
\end{eqnarray}
where ${}_{n}C_{m}=n!/(n-m)!m!$.
With this general probability function,
we can evaluate the expectation value for the number of peaks found in the noise field with
a given S/N threshold:
\begin{eqnarray}
	N = \int_{(S/N){\rm th}}^{\infty} \sum_{m=0}^{n-1}P_{n}^{(m)}(x)dx
\end{eqnarray}
According to this formula, we should find on average 2.7 and 25.3  spurious peaks exceeding $(S/N)_{\rm th}=4.5$ and $4.0$, respectively,
for a 1000 deg$^{2}$ survey with 1.5$^{\prime}$ Gaussian smoothing.

\begin{figure}[htbp] 
   \centering
   \includegraphics[width=6in]{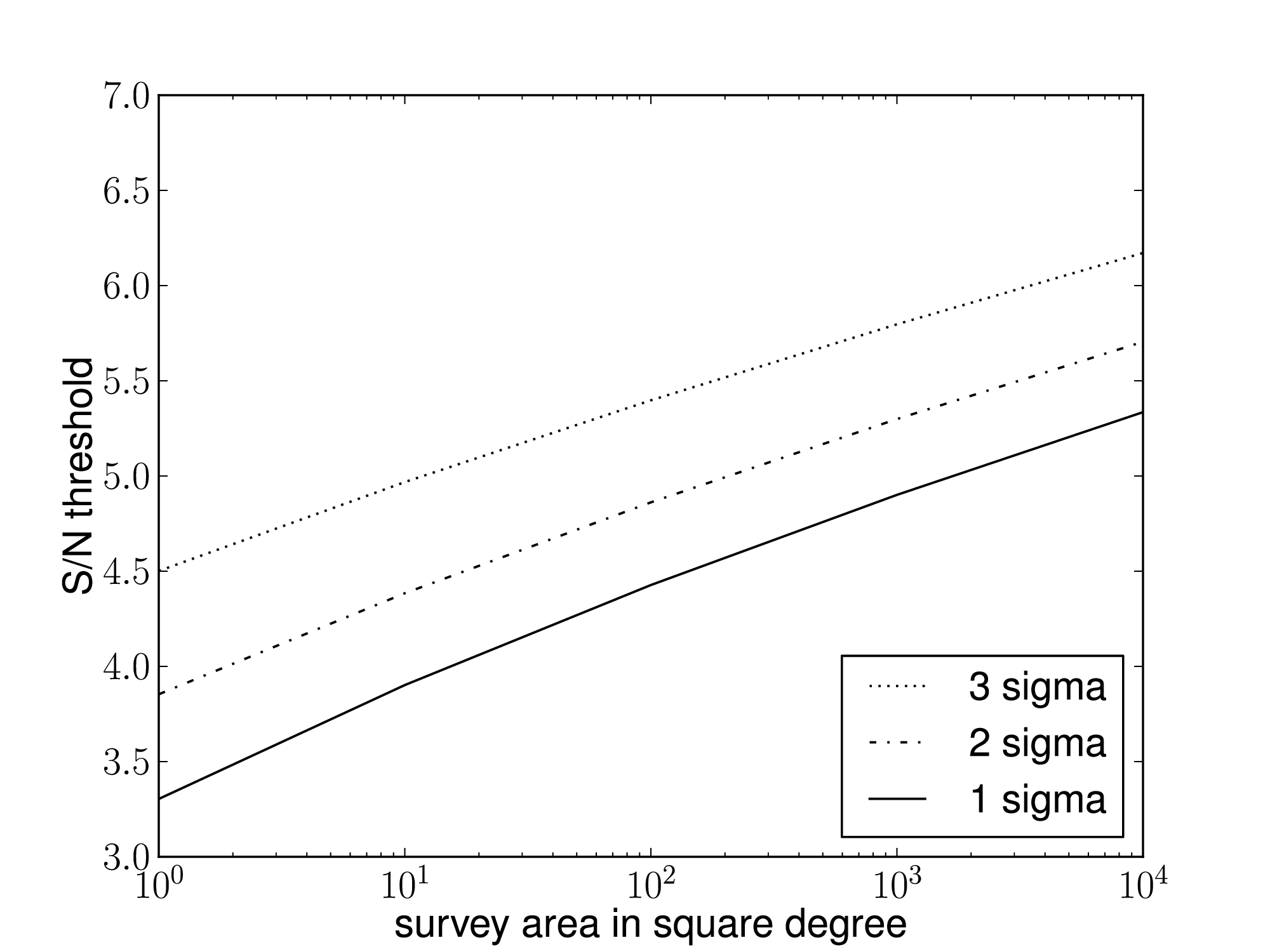} 
   \caption{S/N threshold dependence on  survey area.}
   \label{fig:SNarea}
\end{figure}

In conclusion, the
analytical expression we derive is useful for future error analysis of large surveys.
If we  place a S/N $= 4.5$ threshold on the B-mode-free kappa (S/N) map,
we expect these high significant peaks to be uncontaminated  at the 99.7\%  confidence level for a 1 deg$^{2}$ field;
only 2.7 spurious peaks  should be found on average for a 1000 deg$^{2}$ survey.
Figure \ref{fig:SNarea} shows the S/N threshold as a function of  survey area.

\section{The Nature of the Weak Lensing Peaks}

To test the reliability of the peak catalogs for the GTO (Figure \ref{fig:MassMapGTO1.5}) and DLS fields (Figure\ref{fig:MassMapDLS1.5}),
we next search for corresponding systems of galaxies.
Weak lensing only traces the surface mass density.
Thus our peaks may be single isolated clusters or superpositions of
less massive groups along the line of sight.
Here we test the peaks against dense foreground redshift surveys following 
\citet{2005ApJ...635L.125G} and \citet{2012ApJ...750..168K}.

\subsection{Redshift Surveys}

Dense redshift samples are available for both the GTO field \citep{2012ApJ...750..168K} and the DLS field
\citep{2005ApJ...635L.125G,2010ApJ...709..832G,2012ApJ...758...25H}.
The redshift surveys were carried out with the  Hectospec \citep{1998SPIE.3355..285F,2005PASP..117.1411F},
a 300-fiber robotic instrument mounted on the MMT.
The spectra cover the wavelength range 3,500-10,000\AA\ with a resolution of $\sim 6$ \AA.
The typical error in an individual redshift (normalized by $(1+z)$) is
27 km/s for emission line objects and 37 km/s for absorption line objects.
The surveys include
redshifts for galaxies with $R\lesssim20.6$ in the 4 deg$^{2}$ DLS field and 
redshifts for galaxies with  $r_{\rm petro}<21.3$, $r-i>0.4$ and $g-r>1.0$ in the GTO 2deg$^{2}$ field.
With these samples, we can identify massive clusters of galaxies with redshift $z\lesssim0.6$.
The surface number densities of the spectroscopic samples are 0.6 and 1.1 ${\rm arcmin}^{-2}$ for the GTO  and the DLS fields, respectively.

To identify potential galaxy systems associated with the detected weak lensing peaks,
we examine the redshift distributions in  cones of  3$^{\prime}$ and 6$^{\prime}$ radius centered on the location of each of the  weak lensing peaks. 

We examine peaks in the redshift histograms. We define the significance of each peak as
the difference between the number of galaxies $N$ and mean number of galaxies $\langle N \rangle$
expected in the cone at that redshift divided
by the standard deviation $\sigma$: $(N-\langle N \rangle)/\sigma$.
For neighbors satisfying  $|\Delta z| < 0.004 (1+z) $,	 
we calculate a velocity dispersion 
from the square root of the unbiased variance.
We use the bootstrap to measure the error in the velocity dispersion.

Tables \ref{tab:peaklist} and \ref{tab:peaklistDLS} list the weak lensing peaks for the GTO and DLS fields, respectively. The Tables list the lensing peak rank and significance, the celestial coordinates of the peaks, the rest frame velocity dispersion and its error, the number of peak members and its significance, and the mean redshift. The Table also include additional information from the
literature.

\TWOFIGs{GTO}{00}
{Redshift distribution for Peak GTO 00. The filled histogram shows objects within a cone of 3$^{\prime}$ radius centered  on the weak lensing peak; the open histogram shows objects within a 6$^{\prime}$ cone. Bins in redshift are $0.002(1+z)$ wide.}
{Close-up view for Peak GTO 00. The image is 6$^{\prime}\times 6^{\prime}$. Contours show the weak lensing peak. Small yellow circles identify galaxies in the redshift peak centered at a mean $z = 0.416$.}

\TWOFIGs{GTO}{01}
{Same as Figure \ref{fig:SpecHistogramGTO00} but for Peak GTO01.}
{Close up view for Peak GTO01.}

\TWOFIGs{GTO}{02}
{Same as Figure \ref{fig:SpecHistogramGTO00} but for Peak GTO02.}
{Same as Figure \ref{fig:SpecDistributionGTO00}. The mean redshift of the peak in Figure \ref{fig:SpecHistogramGTO02} corresponding to the weak lensing system is $z = 0.540$ }

\TWOFIGs{GTO}{03}
{Same as Figure \ref{fig:SpecHistogramGTO00} but for Peak GTO03.}
{Same as Figure \ref{fig:SpecDistributionGTO00}. The significant peak in the redshift histogram (Figure \ref{fig:SpecHistogramGTO03}) corresponding to the
lensing peak is $z = 0.731$.}

\TWOFIGs{DLS}{00}
{Same as Figure \ref{fig:SpecHistogramGTO00} but for Peak DLS00}
{Same as Figure \ref{fig:SpecDistributionGTO00}. This weak lensing peak corresponds to a superposition of structures in the redshfit survey at redshifts $z = 0.319$ (yellow circles) and $z = 0.535$ (skyblue circles).}

\TWOFIGs{DLS}{01}
{Same as Figure \ref{fig:SpecHistogramGTO00} but for Peak DLS01}
{Same as Figure \ref{fig:SpecDistributionGTO00}. The mean redshfit of the peak in Figure \ref{fig:SpecHistogramDLS01} corresponding to the weak lensing system is $z = 0.531$.}

\TWOFIGs{DLS}{02}
{Same as Figure \ref{fig:SpecHistogramGTO00} but for Peak DLS02}
{Same as Figure \ref{fig:SpecDistributionGTO00}. The mean redshfit of the peak in Figure \ref{fig:SpecHistogramDLS02} corresponding to the weak lensing system is $z = 0.647$.}

\subsection{The GTO Field}\label{sec:verification}

We list peaks in Table \ref{tab:peaklist} that have S/N $\geq$ 3.7  as in  \cite{2007ApJ...669..714M}.
However, we concentrate on peaks with S/N$\geq$ 4.56 because our analysis of the noise maps shows that this higher threshold yields a more robust catalog.

The most significant peak, 0, has an obvious counterpart 
in the redshift histogram at a mean redshift $z = 0.419$ (Figure \ref{fig:SpecHistogramGTO00}).
Figure \ref{fig:SpecDistributionGTO00} shows 
the spatial distribution of
cluster members within $\Delta z = 0.004$ of the mean and within 3${}^{\prime}$ of the weak lensing peak. It is clear that these galaxies  are concentrated around the lensing peak.
There is also extended x-ray emission associated with this system;
the SDSS red sequence survey \citep{2009ApJS..183..197W} also identifies it.

The lensing peak is slightly offset from the galaxy distribution.
Assuming the brightest cluster member is the BCG, its offset is $\sim 1^{\prime}$,
comparable with smoothing length.

Figure \ref{fig:twofigGTO01}, corresponding to peak 1, shows 
a concentration of bright galaxies, but the redshift histogram does not show an obvious peak.
This discrepancy may result from under-sampling at this location in the redshift survey
(see peak 7 in Figure 2 of \citet{2012ApJ...750..168K}). 
\citet{1986ApJ...306...30G} reported that this concentration of galaxies is a cluster with $z=0.2830$.
The redshift histogram does show some galaxies around this redshift.

Peak 2 has a significance of 4.92 (Figure \ref{fig:twofigGTO02}).
This peak has a clear counterpart
in the redshift histogram at a mean $ z = 0.540$. X-ray observations also show extended emission.
Figure \ref{fig:SpecDistributionGTO02} shows the distribution of the suspected cluster members on the sky. 
The lensing peak is shifted by several arcminutes from the peak of galaxy distribution.
The reason for the shift is not clear, but we suspect that because the peak
is near the edge of the imaging field,  there may be distortion in the $\kappa$-S/N map. 
This peak may also  be affected by a neighboring bright star.

Peak 3 has a significance of 4.76 and, remarkably, corresponds to a peak in the redshift survey
at $ z = 0.73$ (Figure \ref{fig:twofigGTO02}).  
Although the spectroscopic sample is small, 
there is an excess in the galaxy count within 3$^{\prime}$ of the peak.

Above the S/N = 4.56 threshold, all of the peaks (with the possible exception of peak 1) correspond to systems of galaxies. We suspect that the failure to detect peak 1 in the redshift survey results from undersampling. The offsets in position between the lensing peaks and the lenses are consistent with the
smoothing length except in a case where edge effects may be important.

Below the S/N = 4.56 threshold many fewer peaks correspond to systems of galaxies as we expect based on analysis of the noise maps. Table \ref{tab:peaklist} lists possible systems associated with 
S/N$\geq 3.7$ peaks for comparison with previous work.

\renewcommand{\thefootnote}{\fnsymbol{footnote}}
\begin{table*}[thdp]
\rotatebox[]{90}{
\begin{minipage}[]{\textheight}
\caption[Peak list for the GTO field.]{
Peak list for the GTO field.
}\label{tab:peaklist}
\centering
\begin{tabular}{c|c|c|c|ccc|ccc|c|cc}
\hline
Rank	&	$\nu$	&	RA$_{2000}$	&	DEC$_{2000}$	&	$\sigma_{p,6}$[km/s]	&	$N_{6}$	&	$z_{6}$	&	$\sigma_{p,3}$[km/s]	&	$N_{3}$	&	$z_{3}$	&	NED	&	X-ray	&	distance\\
\hline 
0	&	5.95	&	16:02:50.826	&	+43:35:51.34	&	$992.0\pm94.5$	&	$40(8.6\sigma)$	&	$0.415$(\#1)		&	$1014.1\pm113.2$	&	$24(11.8\sigma)$	&	$0.416$(\#1)	&	$0.420$\footnotemark[2], XrayS\footnotemark[4]  &	$9.10\pm1.32$\footnotemark[6]	&	1.156'\\
1	&	5.07	&	16:07:58.418	&	+43:38:23.99	&	$\bullet$\footnotemark[1]			&					&					&	$\bullet$\footnotemark[1]			&									&				&	$0.2830$\footnotemark[3] &	&	\\
2	&	4.94	&	16:03:01.380	&	+42:46:32.00	&	$655.8\pm94.8$	&	$29(12.0\sigma)$	&	$0.540$(\#0)		&	$728.3\pm127.7$	&	$18(13.9\sigma)$	&	$0.540$(\#0)	&	$0.556$\footnotemark[2],$0.5391$\footnotemark[3],XrayS\footnotemark[4]  &	$20.0\pm1.5$\footnotemark[6]	&	1.931'\\
3	&	4.76	&	16:05:24.866	&	+42:45:34.17	&	$497.9\pm136.5$	&	$10(16.1\sigma)$	&	$0.731$			&	$562.8\pm151.1$	&	$8(26.2\sigma)$	&	$0.731$		&	 &	&	\\
\hline
4	&	4.38	&	16:05:39.982	&	+43:22:13.70	&					&					&					&					&					&					&	 				&					&	\\
5	&	4.29	&	16:04:37.615	&	+43:27:34.65	&					&					&					&					&					&					& XrayS\footnotemark[5] 			&					&	\\
6	&	4.19	&	16:04:41.282	&	+42:37:54.73	&					&					&					&					&					&					& 					&					&	\\
7	&	4.17	&	16:07:44.303	&	+42:37:45.58	&					&					&					&					&					&					&					&					&	\\
8	&	4.12	&	16:04:10.464	&	+42:39:14.47	&	$411.9\pm74.9$	&	$9(18.1\sigma)$	&	$0.673$(\#3)		&	$508.6\pm102.4$	&	$6(28.1\sigma)$	&	$0.673$(\#3)		&	 				&					&	\\
9	&	3.85	&	16:05:15.727	&	+42:39:14.39	&					&					&					&					&					&					& 					&					&	\\
10	&	3.70	&	16:01:46.456	&	+42:56:06.40	&	$421.9\pm59.1$	&	$15(4.5\sigma)$	&	$0.256$			&	$366.1\pm59.7$	&	$10(7.2\sigma)$	&	$0.287$			&					&					&	\\
\hline
\hline
\end{tabular}
\footnotetext[0]{The column "Rank" is sorted by the lensing significance $\nu$. $\sigma_{p}$ shows a measured velocity dispersion, a number of galaxies on the cone and its mean redshift for 6 and 3 arcmin cones, respectively.
NED: objects classified as ``Galaxy Clusters'' or ``X-ray sources (XrayS)'' in NED within 3 arcmin,
The column "X-ray" and distance shows a flux of X-ray in the literature and a positional difference between detected peak and X-ray source.}
\footnotetext[1]{This peak is not well sampled at this location in the redshift survey}
\footnotetext[2]{WHO: \citep{2009ApJS..183..197W, 2010ApJS..187..272W},}
\footnotetext[3]{GHO: \citep{1986ApJ...306...30G},}
\footnotetext[4]{1WGA: ftp://cdsarc.u-strasbg.fr/cats/IX/30/ReadMe,}
\footnotetext[5]{KLG2009: \citep{2009ApJ...690..295K}}
\footnotetext[6]{X-ray flux is converted from count rate in 0.24-2.0 keV with a constant correction factor of $1.5\times10^{-11} {\rm erg}/{\rm cm}^2/{\rm sec}$ in the unit of $10^{-14} {\rm erg}/{\rm cm}^2/{\rm sec}$ \citep{2000yCat.9031....0W} .}
\end{minipage}
}
\end{table*}%

\subsection {The DLS Field}

In the DLS field (Figure \ref{fig:MassMapDLS1.5}),
the most significant peak (0) has S/N = 6.18 (Figure \ref{fig:twofigDLS00}).
The redshift histogram 
(Figure \ref{fig:SpecDistributionDLS00}) shows two different 
concentrations centered at $z=0.320$ and $z=0.537$.
This peak is apparently a superposition of clusters along the line of sight.
So far, this peak is not covered by X-ray observations.

Peak 1 (Figure \ref{fig:twofigDLS01}) has S/N = 4.98 but it
unfortunately coincides with bright stars.
The center of the weak lensing peak is masked by these bright stars.
The redshift histogram shows a peak at $z=0.531$, but the velocity dispersion for the $3^{\prime}$ cone,
$\sigma \sim 386$ km s$^{-1}$ is too low to account for the lensing peak:  the 636 km s$^{-1}$ dispersion for the $6^{\prime}$ cone is sufficient to account for the detection.
The coincident star compromises accurate measurement of the velocity dispersion.

Two other previous studies (lensing and X-ray)
\citep{2009ApJ...702..603A,2009ApJ...702..980K}
reported that there is a cluster at $z = 0.53$.
X-ray observations show  extended emission suggesting that the velocity dispersion may be underestimated. The redshift distribution also shows a peak at
at $z=0.183$
Thus the system at $z = 0.53$ may be boosted by the foreground large-scale structure.

The third significant Peak 02 (Figure \ref{fig:twofigDLS02}) has S/N = 4.56 and coincides with
five galaxies around $z = 0.65$  in the spectroscopic data (Figure \ref{fig:SpecDistributionDLS02})
even though such high $z$ galaxies are  rare in the redshift survey. 
There is also extended x-ray emission associated with this peak.

In the DLS field, the results are less clear than in the GTO field. One of the peaks (2) is a cluster
and one (0) is a clear superposition of clusters. The interpretation of peak 1 is unclear, but we believe the peak is not noise.
In Table \ref{tab:peaklistDLS} we also list lower significance peaks and possibly associated systems of galaxies for comparison with previous work.

\begin{table*}[htdp]
\rotatebox{90}{\begin{minipage}{\textheight}
\caption{Peak list for the DLS field.}
\begin{center}
\begin{tabular}{c|c|c|c|ccc|ccc|c|cc}
\hline
\hline
Rank	&	$\nu$	&	RA$_{2000}$	&	DEC$_{2000}$	&	$\sigma_{p,6}$[km/s]	&	$N_{6}$	&	$z_{6}$	&	$\sigma_{p,3}$[km/s]	&	$N_{3}$	&	$z_{3}$	&	NED	&	X-ray	&	distance\\
\hline 
0	&	6.18	&	09:16:52.817	&	+30:00:28.44	&	$497\pm50$	&	$47(7.8\sigma)$	&	$0.318$	&	$468\pm61$	&	$23(9.3\sigma)$	&	$0.319$	&	 &		&	\\
	&		&				&				&	$486\pm75$	&	$25(7.0\sigma)$	&	$0.535$	&	$446\pm79$	&	$18(13.7\sigma)$	&	$0.535$	&	 &		&	\\
1	&	4.98	&	09:15:54.449	&	+29:37:08.25	&	$636\pm133\footnotemark[1]$	&	$26(7.2\sigma)$	&	$0.531$ 	&	$386\pm64\footnotemark[1]$	&	$12(9.2\sigma)$	&	$0.531$ 	&	 $0.53$$\footnotemark[2]\footnotemark[3]$	&	$4.99\pm1.32(2.97)$\footnotemark[4]	&	0.797'\\
2	&	4.65	&	09:16:04.967	&	+30:28:08.30	&	$524\pm145$	&	$5(30.2\sigma)$	&	$0.647$	&	$524\pm147$	&	$5(54.1\sigma)$		&	$0.647$	&	 &	$9.52\pm2.24(4.60)$	&	0.786'\\
	&		&				&				&				&					&			&				&					&			&	&	$1.75\pm1.21(0.65)$\footnotemark[4]	&	1.677'\\
\hline
3	&	4.46	&	09:15:54.292	&	+30:03:08.24	&				&					&			&				&					&			&	&		&	\\
4	&	4.24	&	09:17:12.960	&	+30:18:28.09	&	$811\pm124$	&	$19(5.3\sigma)$	&	$0.535$		&	$859\pm162$	&	$12(9.1\sigma)$	&	$0.535$		&	 &	$29.1\pm10.4(5.88)$\footnotemark[4] &	0.961'\\
5	&	3.86	&	09:16:20.559	&	+29:15:48.60	&	$1026\pm188$	&	$13(11.2\sigma)$	&	$0.365$		&	$486\pm187$	&	$8(17.7\sigma)$	&	$0.367$		&	 &		&	\\
	&		&				&				&	$818\pm99$	&	$25(7.0\sigma)$	&	$0.536$		&	$968\pm169$	&	$12(9.1\sigma)$	&	$0.536$		&	 &		&	\\
\hline
\hline
\end{tabular}
\end{center}
\label{tab:peaklistDLS}
\footnotetext[0]{Same as Table \ref{tab:peaklist} but Peak list for the DLS field. NED: objects classified as ``Galaxy Clusters'' or ``X-ray sources (XrayS)'' in NED within 3 arcmin}
\footnotetext[1]{This peak coincides a foreground bright star and then prevent us to measure accurate velocity dispersion.}
\footnotetext[2]{AWM2009 \citep{2009ApJ...702..603A}}
\footnotetext[3]{KKD2009 \citep{2009ApJ...702..980K}.}
\footnotetext[4]{X-ray flux in 0.2keV-12keV(0.2keV-2keV) band for extended soruce (ext$>0$ arcsec) in the unit of $10^{-14}{\rm erg}/{\rm cm}^2/{\rm s}$ \citep{2010yCat.9041....0X} }
\end{minipage}}
\end{table*}%

\section{Comparison with Earlier Results}

The new maps of the GTO field and our 1 deg$^2$ portion of the DLS field contain seven weak lensing peaks above our threshold; these peaks are probably all associated either with a single system of galaxies or with a superposition along the line-of-sight. In other words, the analysis suggests that the lensing signal above our threshold  is a reasonably robust measure of features of the large-scale structure of the universe.

Next we compare the results for the lensing peaks with previous work. The issues relevant for the GTO and DLS field are slightly different. In the GTO case, we analyze the same data as \cite{2012ApJ...750..168K}, but we apply new procedures to reduce systematic error. In the DLS subfield, we apply essentially the same reduction procedure as the GTO field. Our Subaru DLS data are deeper and are taken in better conditions than the original data analyzed by \citet{2006ApJ...643..128W}.

For the GTO field,  \cite{2012ApJ...750..168K}
find two peaks with S/N $\geq 4.5$; we also find these peaks at S/N $\geq 4.56$ along with two additional ones.
This difference occurs because the suppression of systematic errors changes the local 
peak values in the $\kappa$-S/N map. There is no change in the average sensitivity of the map.

The difference in the peak height distribution for the B-mode map of
\citet{2012ApJ...750..168K} and our map demonstrates again that the systematic error suppression is effective.
\citet{2012ApJ...750..168K} found 4 peaks with S/N$\geq 3.7$ in their B-mode map.
We find only 2 peaks with this amplitude in our new B-mode map  (S/N=3.90, 3.87). 
The highest B-mode peak now has S/N = 3.90 rather than 4.23. 

Our data cover about 25\% of the original DLS field F2. 
\cite{2006ApJ...643..128W} identified one peak and 
\cite{2009ApJ...702..980K} detected 3 peaks in this region of F2.
All of these peaks have S/N$ < 3.9$, substantially below our revised threshold.

The original DLS F2 data were taken on the Kitt Peak 4-meter in the R-band image with PSF $<0.9^{\prime\prime}$;
final co-added images reached a depth of $R\sim26$
and the source galaxy density is, $\sim 20~{\rm arcmin}^{-2}$.
The Subaru source density is $\sim 30~ {\rm arcmin}^{-2}$ to roughly the same magnitude limit.
The lower source galaxy density may result from the poorer seeing at Kitt Peak. In this case, the Subaru map is more sensitive mainly as a result of the increase source density. We find three peaks above the threshold S/N = 4.56; one of these is a cluster at $z = 0.64$.

Obviously, at S/N below our revised threshold, there are peaks that correspond to systems of galaxies (Tables \ref{tab:peaklist}, \ref{tab:peaklistDLS}). As we would expect, the fraction of ``real'' peaks declines with decreasing S/N.

\section {Conclusion}

Weak lensing is a powerful tool of modern cosmology. Its application to the detection of clusters of galaxies is limited to some extent by subtle systematic errors in the construction of the $\kappa$-S/N maps. Careful analysis of the B-mode maps where there should be no signal provides a route to
reduction of these systematic errors. We use a set of noise maps along with the B-mode maps for two fields observed with the Subaru telescope to test procedures for the reduction of systematics.

Our tests of the B-mode maps against noise maps demonstrate that the following image processing procedures significantly reduce systematic error in the final $\kappa$-S/N map:
\begin{itemize}
\item Registration of positions onto the absolute celestial coordinate grid,
\item Simultaneous application of all image transformations to avoid loss of information in interpolation, and
\item Application of a smoothing cutoff at 10$^\prime$ to remove excess large-scale power.
\end{itemize}
By comparing the S/N distribution for noise and B-mode maps, we demonstrate that these revised procedures significantly reduce the systematics that otherwise appear in the B-mode map. We then take advantage of these revised procedures to show that weak lensing cluster
detection is insensitive to the smoothing length for the $\kappa$-S/N map in the range $1.0^{\prime}$ to 1.5$^{\prime}$. Also on the basis of this analysis, we suggest use of the 99\% threshold derived from the ``most extreme peaks'' among the highest peaks of the noise maps rather than the more standard 3$\sigma$ threshold. We provide an analytic expression for the distribution of these highest peaks that can be used to estimate the fraction of false peaks in the $\kappa$-S/N maps as a function of the detection threshold.  

We test our systematic error reduction procedure and our high peak analysis by analyzing the GTO field and  a portion of the DLS F2 field; together these fields cover $2.52 {\rm deg}^2$.
A threshold of 4.56 corresponds to the 99.5\% confidence level from the highest peak analysis
for our 1.5$^{\prime}$ smoothing length. We expect these peaks to be uncontaminated by false peaks over this total area. The data show seven peaks above this threshold; all of these peaks correspond to galaxy overdensities. Six of these peaks correspond to clusters of galaxies; one is a superposition of systems. Five of these peaks are well-sampled by dense foreground redshift surveys providing reliable estimates of the rest frame line-of-sight velocity dispersions of the clusters. One peak in the GTO field is in region sparsely sampled by the redshift survey, but a concentration of galaxies is evident in the imaging data. For one peak in the DLS field, foreground stars probably lead to an underestimate of the velocity dispersion from the spectroscopic data, but the peak is associated with extended x-ray emission. Taken together, these results substantiate the efficacy of our revised procedures.

The systematic error reduction procedures we apply are general and can be applied to future large-area weak lensing surveys. Our high peak analysis suggests that with a S/N threshold of 4.5, there should be only 2.7 spurious weak lensing peaks even in an area of 1000 deg$^2$ where we expect $\sim$ 2000
peaks based on our Subaru fields.

\acknowledgments


Masafumi Yagi provided a technique for analytical treatment of the maximum value distribution.
Also we acknowledge useful discussions with Satoshi Kawanomoto, Hisanori Furusawa and Yutaka Komiyama about Suprime-Cam. We thank the MMT remote observers Perry Berlind and Mike Calkins for operating the Hectospec and we thank Susan Tokarz for reducing the Hectospec data.

Yousuke Uutsumi acknowledges financial support from
the Japan Society for the Promotion of Science (JSPS) through JSPS
Research Fellowships for Young Scientists
and from the Department of Astronomical Sciences of the Graduate University for Advanced Studies (SOKENDAI) through Research incentive.

The Smithsonian Institution supports the research of Margaret Geller, Daniel Fabricant, and Michael Kurtz. 

This work was also supported in part by the FIRST program ``Subaru
Measurements of Images and Redshifts (SuMIRe)'', World Premier
International Research Center Initiative (WPI Initiative), MEXT,
Japan, and Grant-in-Aid for Scientific Research from the JSPS
(23740161).

The authors wish to recognize and acknowledge the very significant
cultural role and reverence that the summit of Mauna Kea has always
had within the indigenous Hawaiian community.  We are most fortunate
to have the opportunity to conduct observations from this sacred
mountain.

{\it Facilities:}\facility{Subaru(Suprime-Cam)}, \facility {MMT(Hectospec)}
\bibliographystyle{apj}
\bibliography{References}

\appendix

\section{Masks}\label{appendix:mask}
Here we present details of masks described in Section \ref{galaxycatalog}.
Table \ref{tab:mask} shows the mask for brights stars based on the external catalog USNO-A2.0 \citep{1998AAS...19312003M}.
Table \ref{tab:mask:GTO} and Table \ref{tab:mask:DLS} show the coordinates for the additional manually constructed masks.

\begin{table}[htdp]
\caption{Masking Radius Around Bright Stars}
\begin{center}
\begin{tabular}{c|c}
\hline
Magnitude Range &	Mask Radius (arcsec)	\\
\hline
\hline
$R < 9.0$	&	210	\\
$  9.0 < R < 10.0$	&	140\\
$10.0 < R < 12.0$	&	70\\
$12.0 < R < 14.0$	&	35\\
$14.0 < R < 16.0$	&	18\\
$16.0 < R < 18.0$	&	6\\
\hline
\end{tabular}
\end{center}
\label{tab:mask}
\end{table}%

\begin{table}[htdp]
\caption{Manually Masked Regions around Bright Stars in the GTO Field.}
\begin{center}
\begin{tabular}{l|l|r|l}
\hline
RA$_{2000}$	&	DEC$_{2000}$	&	radius	(arcsec)	& Origin\\
\hline
\hline
240.11876	&	42.9889	&	270	&	Filter reflection\\
240.327	&	42.7408	&	270	&	Filter reflection\\
240.44033	&	42.842131	&	220	&	Filter reflection\\
240.45127	&	43.480879	&	270	&	Filter reflection\\
240.48183	&	43.629343	&	41.4	&	Incorrect position\\
240.77826	&	42.616311	&	270	& 	Filter reflection\\
241.06437	&	43.414799	&	99	&	Dewar window reflection\\
241.16981	&	43.42311	&	36	&	Incorrect position\\
241.26726	&	43.561169	&	99	&	Dewar window reflection\\
241.3551	&	43.711462	&	270	&	Filter reflection\\
\hline
\end{tabular}
\end{center}
\label{tab:mask:GTO}
\end{table}%

\begin{table}[htdp]
\caption{Manually Masked Regions Around Bright Stars in the DLS Field.}
\begin{center}
\begin{tabular}{l|l|r|l}
\hline
RA$_{2000}$	&	DEC$_{2000}$	&	radius	(arcsec)	& Origin\\
\hline
\hline
138.888195	&	29.792573	&	120	&	Dewar window reflection\\
139.343445	&	29.452575	&	120	&	Dewar window reflection\\
139.100387	&	30.040381	&	120	&	Dewar window reflection\\
139.254212	&	30.374562	&	120	&	Dewar window reflection\\
139.412442	&	30.320542	&	120	&	Dewar window reflection\\
\hline
\end{tabular}
\end{center}
\label{tab:mask:DLS}
\end{table}%

\end{document}